# Surface effects on anti-plane shear waves propagating in magneto-electro-elastic nano-plates

Bin Wu[1], Chunli Zhang[1], Weiqiu Chen[1,2], Chuanzeng Zhang[3,*]

[1] Department of Engineering Mechanics, Zhejiang University, Hangzhou 310027, China

[2] State Key Lab of CAD & CG, Zhejiang University, Hangzhou 310058, China

[3] Department of Civil Engineering, University of Siegen, Siegen D-57068, Germany

**Abtract:**

Material surface may have a remarkable effect on the mechanical behavior of magneto-electro-elastic (or multiferroic) structures at nano-scale. In this paper, a surface magneto-electro-elasticity theory (or effective boundary condition formulation), which governs the motion of the material surface of magneto-electro-elastic nano-plates, is established by employing the state-space formalism. The properties of anti-plane shear (SH) waves propagating in a transversely isotropic magneto-electro-elastic plate with nano-thickness are investigated by taking surface effects into account. The size-dependent dispersion relations of both antisymmetric and symmetric SH waves are presented. The thickness-shear frequencies and the asymptotic characteristics of the dispersion relations considering surface effects are determined analytically as well. Numerical results show that surface effects play a very pronounced role in elastic wave propagation in magneto-electro-elastic nano-plates, and the dispersion properties depend strongly on the chosen surface material parameters of magneto-electro-elastic nano-plates. As a consequence, it is possible to modulate the waves in magneto-electro-elastic nano-plates through surface engineering.

---

[*] Corresponding author. Tel.: +49 271 7402173; fax: +49 271 7404073.

E-mail address: c.zhang@uni-siegen.de (Ch. Zhang).

# 1. Introduction

In general, multiferroic materials are defined to possess at least two coexisting orders among the electric, magnetic and elastic ones. In particular, however, the remarkable property of multiferroic materials is regarded to be the coupling interaction between magnetic and electric fields, which is known as the magnetoelectric (ME) effect [1]. The ME effect in multiferroic materials has many potential innovative applications in multifunctional devices such as magnetic sensors, current sensors, magnetic energy harvesters, and so on [2-5], and has been studied extensively in many scientific fields. The ME effect in single-phase materials has no value in practice owing to its weakness at room temperature. The existing experimental and theoretical studies [6-8] show that multiferroic composites, especially laminated composites, made of piezoelectric and piezomagnetic or magnetostrictive materials can achieve much larger ME coupling than those in single-phase materials. Therefore, there are increasing research interests in the ME coupling behavior in multiferroic composites in recent years. In addition, researchers have carried out a series of in-depth studies on the static and dynamic responses of magneto-electro-elastic materials/structures within the framework of continuum mechanics [9-11].

Materials at nano-scale have unique optical, electronic, or mechanical properties. With rapid development of nanotechnology, nano-materials/structures, which now can be easily manufactured, have drawn tremendous attention due to their promising applications in future nano-devices [12]. For example, Zheng et al. [13] firstly reported multiferroic nano-films composed of $CoFe_2O_4$-$BaTiO_3$ via physical deposition technique and chemical solution processing. Xie et al. [14], using sol-gel process and electrospinning technique, synthesized multiferroic nano-fibers. The strong ME coupling has been found in these multiferroic composites at nano-scale [13, 14]. Just as elastic and piezoelectric nano-structures, one distinct feature of multiferroic nano-composites is the size-dependent characteristic due to the increasing ratio of surface area to volume. Various size-dependent phenomena have been reported and also well explained by the atomic simulation, the first principle, or the modified continuum mechanics for elastic and piezoelectric nano-structures. One reasonable and also successful explanation is due to the surface effects at nano-scale, which account for the difference between the properties of the bulk and its surface [15-17]. Therefore, the modified continuum mechanics model for multiferroic nano-materials, which takes account of the surface effects, is very vital to better understand and accurately predict the static and dynamic behaviors of multiferroic nano-composites.

In order to analyze the surface effects in elastic nano-materials, a rigorous nonlinear framework of surface elasticity, which was proposed by Gurtin and Murdoch for a deformable material surface [18] and now known as the GM theory, has been extensively adopted in recent years in the study of size-dependent properties and responses of nano-sized materials and structures [19-22]. Steigmann and Ogden [23,24] generalized the GM theory to incorporate the flexural stiffness of the free surface into the surface constitutive equation. Based on Steigmann and Ogden's works [23,24], the dynamic responses of elastic solids with intrinsic boundary elasticity have been studied [25-28]. A further development of the surface elasticity theory can be found in [29,30], where a new energy functional was introduced to propose a hyperelastic surface model based on the framework of finite deformation. Furthermore, the surface elasticity theory has been extended to surface piezoelectricity theory by making use of the phenomenological continuum theory which accounts for the linear interplay between electricity and elasticity [31,32]. Based on the surface piezoelectricity theory [31], substantial researches have been carried out to the surface effects on static and dynamic electromechanical properties of piezoelectric nano-structures [33-37]. Recently, based on Mindlin and Tiersten's thin layer model [38,39], Chen [40,41] established a surface piezoelectricity theory by employing the state-space formalism. Making use of Chen's theory [41], the surface effects on wave propagation in piezoelectric or anisotropic nano-structures have been investigated [40-43]. However, for multiferroic nano-composites, the above surface theories will become insufficient since no ME coupling is involved. Up to now, there is very little attention paid to the mechanical property of multiferroic nano-structures except the paper of Fang et al. [44], where interface energy effects on the propagation of anti-plane shear waves in nano-sized cylindrical piezoelectric/piezomagnetic composites are investigated, and the paper of Yu and Zhang [45], where surface effects on the ME response of layered ME composites with nano-scale thickness are studied.

In this paper, we firstly extend our previous works [40,41] on piezoelectric nano-structures to multiferroic nano-structures and present a theory of surface magneto-electro-elasticity (i.e., the effective boundary conditions) for anti-plane problems by means of the state-space formalism. Based on the derived effective boundary conditions, the dispersion relations of anti-plane shear (SH) waves propagating in a transversely isotropic magneto-electro-elastic nano-plate with surface effects are obtained explicitly. The thickness-shear frequencies (i.e., cutoff-frequencies) and the asymptotic characteristics of the dispersion relations for SH wave modes are also obtained analytically. By making a comparison of the approximate solutions with the exact dispersion relations, the validation ranges of the proposed surface magneto-electro-elasticity theory are determined for SH waves. Finally, numerical results are presented in graphical forms to illustrate the dependence of the SH wave dispersion properties on the surface material parameters.

## 2. Basic equations and state space formalism

At nano-scale, it is well known that the atomic structure in or near the surface of a medium is usually endowed with a different circumstance from that in the bulk counterpart, such that the material properties of the surface are distinct from those of the bulk. Accordingly, we consider an infinite transversely isotropic magneto-electro-elastic nano-plate which is polarized along the $x_3$-axis and illustrated in figure 1. The magneto-electro-elastic nano-plate is considered to consist of a bulk layer and two surface layers. The thickness of the bulk layer is $2H$, while the thickness of the surface layers is $h$.

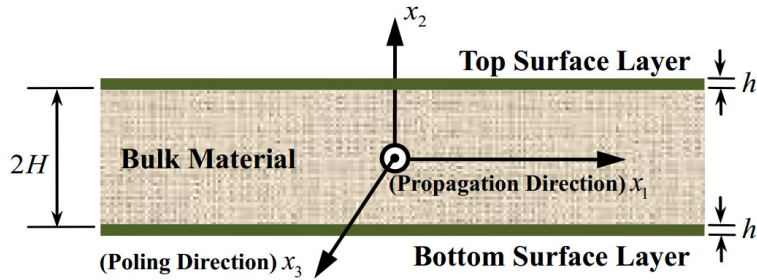

**Figure 1.** A magneto-electro-elastic nano-plate with identical top and bottom surface layers.

For the anti-plane problem which is independent of $x_3$, the displacement components

$u_i$, electric potential $\phi$, and magnetic potential $\psi$ are assumed as

$$u_1 = u_2 = 0, \quad u_3 = u_3(x_1, x_2, t), \quad \phi = \phi(x_1, x_2, t), \quad \psi = \psi(x_1, x_2, t) \tag{1}$$

The differential equations of motion without body forces and the Gaussian equations of magneto-electrostatics without free charges and magnetic induction sources, respectively, are represented by

$$T_{13,1} + T_{23,2} = \rho \ddot{u}_3, \quad D_{1,1} + D_{2,2} = 0, \quad B_{1,1} + B_{2,2} = 0 \tag{2}$$

where the subscript comma stands for the derivative with respect to the spatial coordinate that follows, and the superimposed dot denotes the derivative with respect to time; $T_{ij}$ are the stress components, $\rho$ is the mass density, $D_i$ are the electric displacements, and $B_i$ are the magnetic inductions.

For a linear transversely isotropic magneto-electro-elastic medium with the poling direction along the $x_3$-axis, the constitutive relations for the anti-plane problem are given by

$$\begin{aligned}
&T_{13} = c_{44} u_{3,1} + e_{15} \phi_{,1} + h_{15} \psi_{,1}, \quad T_{23} = c_{44} u_{3,2} + e_{15} \phi_{,2} + h_{15} \psi_{,2}, \\
&D_1 = e_{15} u_{3,1} - \varepsilon_{11} \phi_{,1} - \alpha_{11} \psi_{,1}, \quad D_2 = e_{15} u_{3,2} - \varepsilon_{11} \phi_{,2} - \alpha_{11} \psi_{,2}, \\
&B_1 = h_{15} u_{3,1} - \alpha_{11} \phi_{,1} - \mu_{11} \psi_{,1}, \quad B_2 = h_{15} u_{3,2} - \alpha_{11} \phi_{,2} - \mu_{11} \psi_{,2}, \\
&T_{11} = T_{22} = T_{33} = T_{12} = 0, \quad D_3 = B_3 = 0
\end{aligned} \tag{3}$$

where $c_{44}$, $e_{15}$, $h_{15}$, $\varepsilon_{11}$, $\alpha_{11}$, and $\mu_{11}$ are the elastic constant, piezoelectric constant, piezomagnetic constant, dielectric constant, ME constant, and magnetic permeability, respectively.

If we choose $\mathbf{u} = [u_3, D_2, B_2]^T$ and $\mathbf{T} = [T_{23}, \phi, \psi]^T$ (the superscript T signifies transpose) and combine them into a state vector, we can obtain from (2) and (3) the following state equation

$$\frac{\partial}{\partial x_2} \begin{Bmatrix} \mathbf{u} \\ \mathbf{T} \end{Bmatrix} = \mathbf{A} \begin{Bmatrix} \mathbf{u} \\ \mathbf{T} \end{Bmatrix} \equiv \begin{bmatrix} \mathbf{A}_{11} & \mathbf{A}_{12} \\ \mathbf{A}_{21} & \mathbf{A}_{22} \end{bmatrix} \begin{Bmatrix} \mathbf{u} \\ \mathbf{T} \end{Bmatrix} \tag{4}$$

where $\mathbf{A}$ is the $6 \times 6$ system matrix, with its four partitioned $3 \times 3$ sub-matrices being

$$\mathbf{A}_{11} = \mathbf{A}_{22}^{\mathrm{T}} = \begin{bmatrix} 0 & P_{12} & P_{13} \\ -e_{15}\dfrac{\partial^2}{\partial x_1^2} & 0 & 0 \\ -h_{15}\dfrac{\partial^2}{\partial x_1^2} & 0 & 0 \end{bmatrix}, \quad \mathbf{A}_{12} = \begin{bmatrix} P_{11} & 0 & 0 \\ 0 & \varepsilon_{11}\dfrac{\partial^2}{\partial x_1^2} & \alpha_{11}\dfrac{\partial^2}{\partial x_1^2} \\ 0 & \alpha_{11}\dfrac{\partial^2}{\partial x_1^2} & \mu_{11}\dfrac{\partial^2}{\partial x_1^2} \end{bmatrix},$$

$$\mathbf{A}_{21} = \begin{bmatrix} \rho\dfrac{\partial^2}{\partial t^2} - c_{44}\dfrac{\partial^2}{\partial x_1^2} & 0 & 0 \\ 0 & P_{22} & P_{23} \\ 0 & P_{23} & P_{33} \end{bmatrix} \tag{5}$$

In equation (5),

$$P_{ij} = Q_{ij} / \det(\mathbf{G}), \quad \mathbf{G} = \begin{bmatrix} c_{44} & e_{15} & h_{15} \\ e_{15} & -\varepsilon_{11} & -\alpha_{11} \\ h_{15} & -\alpha_{11} & -\mu_{11} \end{bmatrix} \tag{6}$$

where $Q_{ij}$ are the cofactors of $\mathbf{G}$.

## 3. Surface magneto-electro-elasticity

In this section, we will follow the method proposed by Chen [40,41] to develop the surface magneto-electro-elasticity for magneto-electro-elastic nano-plates. The starting point is the state equation (4). The two surface layers are modeled as thin magneto-electro-elastic material layers with the same material properties and the same thickness $h$. As described above, the material properties of the surface layers can be different from those of the bulk material. Instead of directly treating the surface layer as a different material phase, we will present the so-called effective boundary conditions which govern the motion of the surface layers in an approximate way.

For clarity, a superscript $s$ will be adopted to indicate the quantities that are associated with the surface layers. Applying (4) to the top surface layer $H \leq x_2 \leq H+h$, we get

$$\frac{\partial}{\partial x_2} \begin{Bmatrix} \mathbf{u}^s \\ \mathbf{T}^s \end{Bmatrix} = \mathbf{A}^s \begin{Bmatrix} \mathbf{u}^s \\ \mathbf{T}^s \end{Bmatrix} \tag{7}$$

The material constants in $\mathbf{A}^s$ are assumed to be independent of $x_2$. Therefore, by treating the partial differential operators $\partial/\partial x_1$ and $\partial/\partial t$ as usual parameters, the solution to (7) may be formally written as

$$\begin{Bmatrix} \mathbf{u}^s(x_2) \\ \mathbf{T}^s(x_2) \end{Bmatrix} = \exp[\mathbf{A}^s(x_2 - H)] \begin{Bmatrix} \mathbf{u}^s(H) \\ \mathbf{T}^s(H) \end{Bmatrix} \qquad (8)$$

Setting $x_2 = H + h$ in (8) yields the following transfer relation

$$\begin{Bmatrix} \mathbf{u}^s(H+h) \\ \mathbf{T}^s(H+h) \end{Bmatrix} = \exp(\mathbf{A}^s h) \begin{Bmatrix} \mathbf{u}^s(H) \\ \mathbf{T}^s(H) \end{Bmatrix} \equiv \mathbf{F} \begin{Bmatrix} \mathbf{u}^s(H) \\ \mathbf{T}^s(H) \end{Bmatrix} \qquad (9)$$

where $\mathbf{F} = \exp(\mathbf{A}^s h)$ is the transfer matrix. By the definition of the matrix exponential, we have the following Taylor-series expansion

$$\mathbf{F} = \mathbf{I} + \mathbf{A}^s h + \frac{1}{2}(\mathbf{A}^s)^2 h^2 + \cdots + \frac{1}{n!}(\mathbf{A}^s)^n h^n + O(h^{n+1}) \qquad (10)$$

Substituting (10) into (9), we can obtain the following relations, which are accurate up to the first order $O(h)$, as

$$\begin{Bmatrix} \mathbf{u}^s(H+h) \\ \mathbf{T}^s(H+h) \end{Bmatrix} = (\mathbf{I} + \mathbf{A}^s h) \begin{Bmatrix} \mathbf{u}^s(H) \\ \mathbf{T}^s(H) \end{Bmatrix} \qquad (11)$$

If we consider the traction-free, magnetically open and electrically shorted boundary conditions at $x_2 = H + h$, then we have

$$\mathbf{T}^s(H+h) = \mathbf{0} \qquad (12)$$

On the other hand, the state variables of the top surface layer should be equal to those of the bulk material at $x_2 = H$, i.e.,

$$\mathbf{T}^s(H) = \mathbf{T}^b(H), \quad \mathbf{u}^s(H) = \mathbf{u}^b(H) \qquad (13)$$

where a superscript $b$ denotes the state vector associated with the bulk material. By using (12) and (13), the last three equations in (11) give the $O(h)$ effective boundary conditions for the top magneto-electro-elastic surface layer as

$$\mathbf{T}^b(H) + h[\mathbf{A}^s_{21}\mathbf{u}^b(H) + \mathbf{A}^s_{22}\mathbf{T}^b(H)] = \mathbf{0} \qquad (14)$$

Similarly, for the bottom surface layer $-H - h \leq x_2 \leq -H$, we have

$$\mathbf{T}^b(-H) - h[\mathbf{A}^s_{21}\mathbf{u}^b(-H) + \mathbf{A}^s_{22}\mathbf{T}^b(-H)] = \mathbf{0} \qquad (15)$$

Equations (14) and (15) are the $O(h)$ effective boundary conditions or the first-order surface magneto-electro-elasticity theory for a planar magneto-electro-elastic surface layer. In fact, effective boundary conditions of an arbitrary order governing the magneto-electro-elastic surface layer may also be readily obtained by appropriately truncating the Taylor-series in (10). It should be emphasized again that these effective conditions are derived for the traction-free, magnetically open and electrically shorted boundary conditions in (12). If different boundary conditions are imposed on the outside surface of the boundary layer, we can deduce the corresponding effective boundary conditions similarly [40]. The effective boundary conditions (14) and (15) for the anti-plane problem also can be easily extended to the three-dimensional case [41].

## 4. SH waves in magneto-electro-elastic nano-plates with surface effects

*4.1. Dispersion relations of SH waves*

It is convenient to define two functions $\xi$ and $\zeta$ as [46]

$$\xi \equiv e_{15}u_3 - \varepsilon_{11}\phi - \alpha_{11}\psi, \quad \zeta \equiv h_{15}u_3 - \alpha_{11}\phi - \mu_{11}\psi \tag{16}$$

Then, (2) and (3) can be reduced as

$$c_{44}^*\nabla^2 u_3 = \rho\ddot{u}_3, \quad \nabla^2\xi = 0, \quad \nabla^2\zeta = 0 \tag{17}$$

where $\nabla^2$ and $c_{44}^*$ are the two-dimensional Laplacian operator and magneto-piezoelectrically stiffened elastic constant

$$\nabla^2 \equiv \partial^2/\partial x_1^2 + \partial^2/\partial x_2^2, \quad c_{44}^* = c_{44} + e_{15}m_1 + h_{15}m_2 \tag{18}$$

with $m_1 = P_{12}/P_{11}, m_2 = P_{13}/P_{11}$. Solving (16), the electric and magnetic potentials in terms of $\xi$ and $\zeta$ are given as

$$\phi = m_1 u_3 + b_1\xi + b_2\zeta, \quad \psi = m_2 u_3 + b_2\xi + b_3\zeta \tag{19}$$

where

$$b_1 = \frac{-\mu_{11}}{\mu_{11}\varepsilon_{11} - \alpha_{11}^2}, \quad b_2 = \frac{\alpha_{11}}{\mu_{11}\varepsilon_{11} - \alpha_{11}^2}, \quad b_3 = \frac{-\varepsilon_{11}}{\mu_{11}\varepsilon_{11} - \alpha_{11}^2} \tag{20}$$

By using (19), the non-zero components of the stresses and electric displacements become

$$T_{13} = c_{44}^* u_{3,1} - m_1 \xi_{,1} - m_2 \zeta_{,1}, \quad T_{23} = c_{44}^* u_{3,2} - m_1 \xi_{,2} - m_2 \zeta_{,2},$$
$$D_1 = \xi_{,1}, \quad D_2 = \xi_{,2}, \quad B_1 = \zeta_{,1}, \quad B_2 = \zeta_{,2} \tag{21}$$

To investigate the SH wave motion in the magneto-electro-elastic nano-plate, the wave solution of (17) may be sought in the following form

$$u_3 = [A\sin(\beta x_2) + B\cos(\beta x_2)] \exp[i(kx_1 - \omega t)],$$
$$\xi = [C\sinh(kx_2) + D\cosh(kx_2)] \exp[i(kx_1 - \omega t)], \tag{22}$$
$$\zeta = [E\sinh(kx_2) + F\cosh(kx_2)] \exp[i(kx_1 - \omega t)]$$

where $\omega$ denotes the circular frequency, $\beta$ and $k$ denote the wave numbers in the $x_2$ and $x_1$-directions (i.e. thickness wave number and propagation wave number), respectively. The solution (22) satisfies (17)$_{2,3}$ automatically, while (17)$_1$ requires

$$\beta^2 = \left(\rho\omega^2 - c_{44}^* k^2\right) / c_{44}^* \tag{23}$$

Substitution of (22) into (19) yields the following electric and magnetic potentials

$$\phi = m_1 [A\sin(\beta x_2) + B\cos(\beta x_2)] + (b_1 C + b_2 E)\sinh(kx_2)$$
$$+ (b_1 D + b_2 F)\cosh(kx_2),$$
$$\psi = m_2 [A\sin(\beta x_2) + B\cos(\beta x_2)] + (b_2 C + b_3 E)\sinh(kx_2) \tag{24}$$
$$+ (b_2 D + b_3 F)\cosh(kx_2)$$

where the common factor $\exp[i(kx_1 - \omega t)]$, which appears in all field variables, is omitted here and after. After substituting (22) into (21)$_{2,4,6}$ we can rewrite the stress component $T_{23}$, electric displacement component $D_2$ and magnetic induction component $B_2$ as follows

$$T_{23} = c_{44}^* \beta [A\cos(\beta x_2) - B\sin(\beta x_2)] - (m_1 C + m_2 E)k\cosh(kx_2)$$
$$- (m_1 D + m_2 F)k\sinh(kx_2),$$
$$D_2 = k[C\cosh(kx_2) + D\sinh(kx_2)], \tag{25}$$
$$B_2 = k[E\cosh(kx_2) + F\sinh(kx_2)]$$

For the traction-free, magnetically open and electrically shorted boundary conditions, the $O(h)$ effective boundary conditions (14) and (15) for the top and bottom surface layers, respectively, are

$$T_{23}(H) + h\left[\rho^s \frac{\partial^2 u_3(H)}{\partial t^2} - c_{44}^s \frac{\partial^2 u_3(H)}{\partial x_1^2} - e_{15}^s \frac{\partial^2 \phi(H)}{\partial x_1^2} - h_{15}^s \frac{\partial^2 \psi(H)}{\partial x_1^2}\right] = 0,$$

$$\phi(H) + h\left[P_{12}^s T_{23}(H) + P_{22}^s D_2(H) + P_{23}^s B_2(H)\right] = 0, \quad (26)$$

$$\psi(H) + h\left[P_{13}^s T_{23}(H) + P_{23}^s D_2(H) + P_{33}^s B_2(H)\right] = 0,$$

and

$$T_{23}(-H) + h\left[c_{44}^s \frac{\partial^2 u_3(-H)}{\partial x_1^2} - \rho^s \frac{\partial^2 u_3(-H)}{\partial t^2} + e_{15}^s \frac{\partial^2 \phi(-H)}{\partial x_1^2} + h_{15}^s \frac{\partial^2 \psi(-H)}{\partial x_1^2}\right] = 0,$$

$$\phi(-H) - h\left[P_{12}^s T_{23}(-H) + P_{22}^s D_2(-H) + P_{23}^s B_2(-H)\right] = 0, \quad (27)$$

$$\psi(-H) - h\left[P_{13}^s T_{23}(-H) + P_{23}^s D_2(-H) + P_{33}^s B_2(-H)\right] = 0$$

Inserting (22)$_1$, (24) and (25) into the $O(h)$ effective boundary conditions (26) and (27), adding and subtracting the resulting equations, we can obtain two sets of three homogeneous equations for the constants $A, C, E$ and $B, D, F$ as follows:

$$\mathbf{K}^{anti}[A, C, E]^\mathrm{T} = \mathbf{0}, \quad \mathbf{K}^{sym}[B, D, F]^\mathrm{T} = \mathbf{0} \quad (28)$$

where $K_{ij}^{anti}$ and $K_{ij}^{sym}$, the elements of the matrices $\mathbf{K}^{anti}$ and $\mathbf{K}^{sym}$, are given by

$$K_{11}^{anti} = c_{44}^* \beta \cos(\beta H) + h[(c_{44}^s + e_{15}^s m_1 + h_{15}^s m_2)k^2 - \rho^s \omega^2]\sin(\beta H),$$

$$K_{1p}^{anti} = k[(e_{15}^s b_{p-1} + h_{15}^s b_p)kh\sinh(kH) - m_{p-1}\cosh(kH)],$$

$$K_{p1}^{anti} = m_{p-1}\sin(\beta H) + hP_{1p}^s c_{44}^* \beta \cos(\beta H), \quad (29)$$

$$K_{2p}^{anti} = b_{p-1}\sinh(kH) + (P_{2p}^s - P_{12}^s m_{p-1})kh\cosh(kH),$$

$$K_{3p}^{anti} = b_p\sinh(kH) + (P_{p3}^s - P_{13}^s m_{p-1})kh\cosh(kH),$$

and

$$K_{11}^{sym} = c_{44}^* \beta \sin(\beta H) + h\left[\rho^s \omega^2 - \left(c_{44}^s + e_{15}^s m_1 + h_{15}^s m_2\right)k^2\right]\cos(\beta H),$$

$$K_{1p}^{sym} = k\left[m_{p-1}\sinh(kH) - \left(e_{15}^s b_{p-1} + h_{15}^s b_p\right)kh\cosh(kH)\right],$$

$$K_{p1}^{sym} = m_{p-1}\cos(\beta H) - hP_{1p}^s c_{44}^* \beta \sin(\beta H), \quad (30)$$

$$K_{2p}^{sym} = b_{p-1}\cosh(kH) + \left(P_{2p}^s - P_{12}^s m_{p-1}\right)kh\sinh(kH),$$

$$K_{3p}^{sym} = b_p\cosh(kH) + \left(P_{p3}^s - P_{13}^s m_{p-1}\right)kh\sinh(kH)$$

where the subscript $p$ takes 2 or 3 in (29) and (30). Thus, it is worth noting that from (28), the antisymmetric ($u_3$ contains sines) and symmetric ($u_3$ contains cosines) solutions of the differential equations with respect to $x_2 = 0$ are not coupled. However, the elastic deformation is coupled to the electric and magnetic fields. We notice here that for magneto-electro-elastic plates, a unique anti-plane elastic motion which is characterized by the vanishing of electric and magnetic fields is found by Chen et al. [9]. The uncoupled anti-plane motion reported in Chen et al. [9] is different from the coupled SH wave motion considered in this study. The difference actually comes from different poling directions. The poling direction here is assumed to be perpendicular to the thickness direction, while in [9], it coincides exactly with the thickness direction.

For nontrivial solutions, the determinant of the coefficient matrices in (28) should vanish, giving rise to the following frequency equations:

$$\left|\mathbf{K}^{anti}\right|=0, \quad \left|\mathbf{K}^{sym}\right|=0, \tag{31}$$

which determine the dispersion relations for the antisymmetric and symmetric SH waves, respectively, propagating in the magneto-electro-elastic nano-plate with surface effects.

*4.2. Dimensionless forms of dispersion relations*

For the purpose of analysis and calculations, it is convenient and useful to rewrite the relevant equations in dimensionless forms. The dimensionless propagation wave number $\gamma$, thickness wave number $\eta$ and circular frequency $\Omega$ are defined as

$$\gamma \equiv 2kH/\pi, \quad \eta \equiv 2\beta H/\pi, \quad \Omega \equiv \omega/\bar{\omega}, \quad \bar{\omega}^2 \equiv \pi^2 c_{44}^*/(4H^2\rho) \tag{32}$$

It is worth noting that very few accurate data regarding surface magneto-electro-elastic properties are available in the literature. Although the values of surface parameters can be obtained from molecular dynamics or atomistic calculations, those values are more qualitative than quantitative. In addition, Chen et al. [43] have already pointed out that according to the physical nature of surface layers, it is more direct and convenient to assign the bulk material constants to the surface layer, just in the manner that the effective boundary conditions are established as described in Section 3. As a result, we introduce the following dimensionless quantities:

$$\begin{aligned}
&r_h = h/H, \quad r_\rho = \rho^s/\rho, \quad r_c = c_{ij}^s/c_{ij}, \\
&r_e = e_{ij}^s/e_{ij} = \varepsilon_{ij}^s/\varepsilon_{ij} = h_{ij}^s/h_{ij} = \alpha_{ij}^s/\alpha_{ij} = \mu_{ij}^s/\mu_{ij}, \\
&\bar{A} = A/H, \quad \bar{C} = C/(e_{15}H), \quad \bar{E} = E/(h_{15}H), \\
&\bar{B} = B/H, \quad \bar{D} = D/(e_{15}H), \quad \bar{F} = F/(h_{15}H)
\end{aligned} \tag{33}$$

where $r_h$, $r_\rho$ and $r_c$ denote the thickness ratio, density ratio and elastic stiffness ratio of the surface layers to the bulk layer, respectively. Here, for simplicity, all the elastic stiffness ratios are assumed to be the same. Also, the piezoelectric constant ratio, piezomagnetic constant ratio, dielectric constant ratio, ME constant ratio, and magnetic permeability ratio are assumed to be identical, denoted by $r_e$ in (33).

Therefore, the relation (23) can be written as

$$\eta^2 = \Omega^2 - \gamma^2 \tag{34}$$

Substituting (32) and (33) into (28)-(31), after rearrangement, we finally obtain

$$\bar{\mathbf{K}}^{anti}\left[\bar{A},\bar{C},\bar{E}\right]^\mathrm{T} = \mathbf{0}, \quad \bar{\mathbf{K}}^{sym}\left[\bar{B},\bar{D},\bar{F}\right]^\mathrm{T} = \mathbf{0} \tag{35}$$

$$\left|\bar{\mathbf{K}}^{anti}\right| = 0, \quad \left|\bar{\mathbf{K}}^{sym}\right| = 0 \tag{36}$$

where

$$\begin{aligned}
\bar{K}^{anti}_{11} &= \frac{\pi}{2}\left(a_1\gamma^2 - r_\rho\Omega^2\right)r_h \tan\left(\frac{\pi}{2}\eta\right) + \eta, \\
\bar{K}^{anti}_{12} &= \gamma\left[\frac{\pi}{2}a_2\gamma r_h \tanh(\frac{\pi}{2}\gamma) - a_3\right], \quad \bar{K}^{anti}_{13} = \gamma\left[\frac{\pi}{2}a_4\gamma r_h \tanh(\frac{\pi}{2}\gamma) - a_5\right], \\
\bar{K}^{anti}_{21} &= a_3 \tan\left(\frac{\pi}{2}\eta\right) + \frac{\pi}{2}e_{15}P^s_{12}\eta r_h, \quad \bar{K}^{anti}_{22} = a_6 \tanh\left(\frac{\pi}{2}\gamma\right) + \frac{\pi}{2}a_7\gamma r_h, \\
\bar{K}^{anti}_{23} &= a_8 \tanh\left(\frac{\pi}{2}\gamma\right) + \frac{\pi}{2}a_9\gamma r_h, \quad \bar{K}^{anti}_{31} = a_5 \tan\left(\frac{\pi}{2}\eta\right) + \frac{\pi}{2}h_{15}P^s_{13}\eta r_h, \\
\bar{K}^{anti}_{32} &= a_8 \tanh\left(\frac{\pi}{2}\gamma\right) + \frac{\pi}{2}a_{10}\gamma r_h, \quad \bar{K}^{anti}_{33} = a_{11} \tanh\left(\frac{\pi}{2}\gamma\right) + \frac{\pi}{2}a_{12}\gamma r_h
\end{aligned} \tag{37}$$

and

$$\begin{aligned}
\bar{K}^{sym}_{11} &= \eta \tan\left(\frac{\pi}{2}\eta\right) + \frac{\pi}{2}r_h\left(r_\rho\Omega^2 - a_1\gamma^2\right), \\
\bar{K}^{sym}_{12} &= \gamma\left[a_3 \tanh\left(\frac{\pi}{2}\gamma\right) - \frac{\pi}{2}a_2\gamma r_h\right], \quad \bar{K}^{sym}_{13} = \gamma\left[a_5 \tanh\left(\frac{\pi}{2}\gamma\right) - \frac{\pi}{2}a_4\gamma r_h\right], \\
\bar{K}^{sym}_{21} &= a_3 - \frac{\pi}{2}e_{15}P^s_{12}\eta r_h \tan\left(\frac{\pi}{2}\eta\right), \quad \bar{K}^{sym}_{22} = a_6 + \frac{\pi}{2}a_7\gamma r_h \tanh\left(\frac{\pi}{2}\gamma\right), \\
\bar{K}^{sym}_{23} &= a_8 + \frac{\pi}{2}a_9\gamma r_h \tanh\left(\frac{\pi}{2}\gamma\right), \quad \bar{K}^{sym}_{31} = a_5 - \frac{\pi}{2}h_{15}P^s_{13}\eta r_h \tan\left(\frac{\pi}{2}\eta\right), \\
\bar{K}^{sym}_{32} &= a_8 + \frac{\pi}{2}a_{10}\gamma r_h \tanh\left(\frac{\pi}{2}\gamma\right), \quad \bar{K}^{sym}_{33} = a_{11} + \frac{\pi}{2}a_{12}\gamma r_h \tanh\left(\frac{\pi}{2}\gamma\right)
\end{aligned} \tag{38}$$

where

$$a_1 = \frac{(c_{44}^s + e_{15}^s m_1 + h_{15}^s m_2)}{c_{44}^*}, \quad a_2 = \frac{e_{15}(e_{15}^s b_1 + h_{15}^s b_2)}{c_{44}^*}, \quad a_3 = \frac{e_{15}}{c_{44}^*} m_1,$$

$$a_4 = \frac{h_{15}(e_{15}^s b_2 + h_{15}^s b_3)}{c_{44}^*}, \quad a_5 = \frac{h_{15}}{c_{44}^*} m_2, \quad a_6 = \frac{e_{15}^2 b_1}{c_{44}^*}, \quad a_7 = \frac{e_{15}^2 (P_{22}^s - P_{12}^s m_1)}{c_{44}^*},$$

$$a_8 = \frac{e_{15} h_{15} b_2}{c_{44}^*}, \quad a_9 = \frac{e_{15} h_{15} (P_{23}^s - P_{12}^s m_2)}{c_{44}^*}, \quad a_{10} = \frac{h_{15} e_{15} (P_{23}^s - P_{13}^s m_1)}{c_{44}^*},$$

$$a_{11} = \frac{h_{15}^2 b_3}{c_{44}^*}, \quad a_{12} = \frac{h_{15}^2 (P_{33}^s - P_{13}^s m_2)}{c_{44}^*}$$

(39)

When both the surface effects and the coupling between the electro-elastic field and the magnetic field are absent (i.e., $r_h = r_\rho = r_c = r_e = 0$ and $h_{ij} = \alpha_{ij} = 0$), the dispersion relations in (36) for SH waves in magneto-electro-elastic nano-plates with surface effects are found to be the same as equations (21) and (27) in [47] for piezoelectric plates without surface effects.

*4.3. Analysis of dispersion relations*

It should be mentioned that the dispersion relations (36), which can predict numerous branches of the dispersion spectrum corresponding to different symmetric and antisymmetric modes, are transcendental equations with respect to the wave parameters $\gamma$, $\eta$, and $\Omega$. Therefore, to obtain detailed and precise numerical information, the dispersion relations (36) in general can only be numerically investigated. However, the thickness-shear frequencies corresponding to cutoff-frequencies at $\gamma = 0$ and asymptotic characteristics of the dispersion relations may be determined analytically with much less calculations [48]. These particular properties of the wave dispersion spectrum are also important to better understand the wave propagation behavior.

In the following, both the dimensionless frequency $\Omega$ and the propagation wave number $\gamma$ are assumed to be real and positive. That is to say, we only pay our attention to time-harmonic traveling waves that are not attenuated.

The thickness-shear waves with $u_1 = u_2 = 0, u_3 = u_3(x_2, t)$ are a special case of the general SH waves. Accordingly, in the limit of $\gamma \to 0$, the dispersion relations (36) for the antisymmetric and symmetric SH waves, respectively, reduce to

$$\begin{vmatrix} \dfrac{\pi}{2}r_\rho r_h \Omega^2 \tan\left(\dfrac{\pi}{2}\Omega\right)-\Omega & a_3 & a_5 \\ a_3 \tan\left(\dfrac{\pi}{2}\Omega\right)+\dfrac{\pi}{2}r_h e_{15} P_{12}^s \Omega & \dfrac{\pi}{2}(a_6+a_7 r_h) & \dfrac{\pi}{2}(a_8+a_9 r_h) \\ a_5 \tan\left(\dfrac{\pi}{2}\Omega\right)+\dfrac{\pi}{2}r_h h_{15} P_{13}^s \Omega & \dfrac{\pi}{2}(a_8+a_{10} r_h) & \dfrac{\pi}{2}(a_{11}+a_{12} r_h) \end{vmatrix} = 0, \qquad (40)$$

and

$$\tan\left(\frac{\pi}{2}\Omega\right) = -\frac{\pi}{2} r_h r_\rho \Omega, \qquad (41)$$

which govern the antisymmetric and symmetric thickness-shear wave modes of the magneto-electro-elastic nano-plate with surface effects, respectively. It is known from (40) and (41) that the antisymmetric thickness-shear frequencies are related to $r_h$, $r_\rho$, $r_c$ and $r_e$, while the symmetric thickness-shear frequencies are only related to $r_h$ and $r_\rho$. If we discard the surface effects (i.e., let $r_h = r_\rho = r_c = r_e = 0$), (40) and (41) become

$$\tan\left(\frac{\pi}{2}\Omega\right) = \frac{\pi}{2} \frac{(a_8^2 - a_6 a_{11})\Omega}{a_3(a_3 a_{11} - a_5 a_8) - a_5(a_3 a_8 - a_5 a_6)} \qquad (42)$$

and

$$\Omega \sin\left(\frac{\pi}{2}\Omega\right) = 0 \qquad (43)$$

It is seen that $\Omega = 0$ satisfies both (42) and (43). While it yields a trivial solution for the antisymmetric thickness-shear modes, it, for the symmetric thickness-shear modes, corresponds to a monolithic rigid-body translation along the $x_3$-axis and an equi-electric/magnetic potentials state. When we neglect the coupling between the electro-elastic field and the magnetic field, i.e., let $h_{ij} = \alpha_{ij} = 0$, (42) and (43) reduce to

$$\tan\left(\frac{\pi}{2}\Omega^*\right) = \frac{\pi}{2}\frac{\Omega^*}{\alpha^2}, \quad \Omega^* \sin\left(\frac{\pi}{2}\Omega^*\right) = 0 \qquad (44)$$

with

$$\Omega^* = \omega/\omega^*, \quad (\omega^*)^2 = \pi^2 \bar{c}_{44}/(4H^2 \rho), \quad \bar{c}_{44} = c_{44} + e_{15}^2/\varepsilon_{11}, \quad \alpha^2 = e_{15}^2/(\bar{c}_{44}\varepsilon_{11}) \qquad (45)$$

where $\bar{c}_{44}$ and $\alpha^2$ are the piezoelectrically stiffened elastic constant and electromechanical coupling factor, respectively. Equation (44) gives the dispersion relations for the well-known antisymmetric and symmetric thickness-shear waves in piezoelectric plates without surface effects [47].

The $\Omega-\gamma$ plane can be divided into two distinct regions depending on the nature of $\eta$. For $\gamma<\Omega$ where $\eta^2=\Omega^2-\gamma^2>0$, the dispersion relations of the SH waves with surface effects are governed by (36)-(39). However, for $\gamma>\Omega$ where $\eta^2=\Omega^2-\gamma^2<0$ and $\eta\equiv i\eta'$ is pure imaginary, the trigonometric functions become hyperbolic tangents in view of $\tan(i\eta')=i\tanh(\eta')$. Therefore, (36)-(38) can be rewritten as

$$\left|\tilde{\mathbf{K}}^{anti}\right|=0,$$
$$\tilde{K}_{11}^{anti}=\frac{\pi}{2}\left(a_1\gamma^2-r_\rho\Omega^2\right)r_h\tanh\left(\frac{\pi}{2}\eta'\right)+\eta',$$
$$\tilde{K}_{12}^{anti}=\gamma\left[\frac{\pi}{2}a_2\gamma r_h\tanh\left(\frac{\pi}{2}\gamma\right)-a_3\right],\quad \tilde{K}_{13}^{anti}=\gamma\left[\frac{\pi}{2}a_4\gamma r_h\tanh\left(\frac{\pi}{2}\gamma\right)-a_5\right],$$
$$\tilde{K}_{21}^{anti}=a_3\tanh\left(\frac{\pi}{2}\eta'\right)+\frac{\pi}{2}e_{15}P_{12}^s\eta'r_h,\quad \tilde{K}_{22}^{anti}=a_6\tanh\left(\frac{\pi}{2}\gamma\right)+\frac{\pi}{2}a_7\gamma r_h,$$
$$\tilde{K}_{23}^{anti}=a_8\tanh\left(\frac{\pi}{2}\gamma\right)+\frac{\pi}{2}a_9\gamma r_h,\quad \tilde{K}_{31}^{anti}=a_5\tanh\left(\frac{\pi}{2}\eta'\right)+\frac{\pi}{2}h_{15}P_{13}^s\eta'r_h,$$
$$\tilde{K}_{32}^{anti}=a_8\tanh\left(\frac{\pi}{2}\gamma\right)+\frac{\pi}{2}a_{10}\gamma r_h,\quad \tilde{K}_{33}^{anti}=a_{11}\tanh\left(\frac{\pi}{2}\gamma\right)+\frac{\pi}{2}a_{12}\gamma r_h \tag{46}$$

and

$$\left|\tilde{\mathbf{K}}^{sym}\right|=0,$$
$$\tilde{K}_{11}^{sym}=-\eta'\tanh\left(\frac{\pi}{2}\eta'\right)+\frac{\pi}{2}r_h\left(r_\rho\Omega^2-a_1\gamma^2\right),$$
$$\tilde{K}_{12}^{sym}=\gamma\left[a_3\tanh\left(\frac{\pi}{2}\gamma\right)-\frac{\pi}{2}a_2\gamma r_h\right],\quad \tilde{K}_{13}^{sym}=\gamma\left[a_5\tanh\left(\frac{\pi}{2}\gamma\right)-\frac{\pi}{2}a_4\gamma r_h\right],$$
$$\tilde{K}_{21}^{sym}=a_3+\frac{\pi}{2}e_{15}P_{12}^s\eta'r_h\tanh\left(\frac{\pi}{2}\eta'\right),\quad \tilde{K}_{22}^{sym}=a_6+\frac{\pi}{2}a_7\gamma r_h\tanh\left(\frac{\pi}{2}\gamma\right),$$
$$\tilde{K}_{23}^{sym}=a_8+\frac{\pi}{2}a_9\gamma r_h\tanh\left(\frac{\pi}{2}\gamma\right),\quad \tilde{K}_{31}^{sym}=a_5+\frac{\pi}{2}h_{15}P_{13}^s\eta'r_h\tanh\left(\frac{\pi}{2}\eta'\right),$$
$$\tilde{K}_{32}^{sym}=a_8+\frac{\pi}{2}a_{10}\gamma r_h\tanh\left(\frac{\pi}{2}\gamma\right),\quad \tilde{K}_{33}^{sym}=a_{11}+\frac{\pi}{2}a_{12}\gamma r_h\tanh\left(\frac{\pi}{2}\gamma\right) \tag{47}$$

It should be noted here that, equations (46) and (47) for $\gamma>\Omega$ can be also obtained alternatively when the solution of the anti-plane displacement $u_3$ in (22)$_1$ is replaced by $u_3 = A\sinh(\beta' x_2) + B\cosh(\beta' x_2)$, where $\beta'^2 = (c_{44}^* k^2 - \rho\omega^2)/c_{44}^* > 0$.

It is worth noting that for some surface material parameters ($r_h, r_\rho, r_c, r_e$), the dispersion branches for both antisymmetric and symmetric SH waves may cross the line $\Omega = \gamma$. Therefore, by taking the limit $\Omega \to \gamma$ in (36)$_1$ and (46), we can determine the locations of the crossover points for the antisymmetric SH waves, which correspond to the roots of the following transcendental equation

$$\begin{vmatrix} \frac{\pi^2}{4}(a_1 - r_\rho)\gamma^2 r_h + 1 & \gamma\left[\frac{\pi}{2}a_2\gamma r_h \tanh\left(\frac{\pi}{2}\gamma\right) - a_3\right] & \gamma\left[\frac{\pi}{2}a_4\gamma r_h \tanh\left(\frac{\pi}{2}\gamma\right) - a_5\right] \\ \frac{\pi}{2}(a_3 + e_{15}P_{12}^s r_h) & a_6 \tanh\left(\frac{\pi}{2}\gamma\right) + \frac{\pi}{2}a_7\gamma r_h & a_8 \tanh\left(\frac{\pi}{2}\gamma\right) + \frac{\pi}{2}a_9\gamma r_h \\ \frac{\pi}{2}(a_5 + h_{15}P_{13}^s r_h) & a_8 \tanh\left(\frac{\pi}{2}\gamma\right) + \frac{\pi}{2}a_{10}\gamma r_h & a_{11} \tanh\left(\frac{\pi}{2}\gamma\right) + \frac{\pi}{2}a_{12}\gamma r_h \end{vmatrix} = 0 \quad (48)$$

Similarly, taking the limit $\Omega \to \gamma$ in (36)$_2$ and (47), the locations of the crossover points for the symmetric SH waves are determined by

$$\begin{vmatrix} \frac{\pi}{2}r_h(r_\rho - a_1)\gamma & a_3 \tanh\left(\frac{\pi}{2}\gamma\right) - \frac{\pi}{2}a_2\gamma r_h & a_5 \tanh\left(\frac{\pi}{2}\gamma\right) - \frac{\pi}{2}a_4\gamma r_h \\ a_3 & a_6 + \frac{\pi}{2}a_7\gamma r_h \tanh\left(\frac{\pi}{2}\gamma\right) & a_8 + \frac{\pi}{2}a_9\gamma r_h \tanh\left(\frac{\pi}{2}\gamma\right) \\ a_5 & a_8 + \frac{\pi}{2}a_{10}\gamma r_h \tanh\left(\frac{\pi}{2}\gamma\right) & a_{11} + \frac{\pi}{2}a_{12}\gamma r_h \tanh\left(\frac{\pi}{2}\gamma\right) \end{vmatrix} = 0 \quad (49)$$

The number of the roots of (48) or (49) depends on the surface material parameters. It can be further shown that, when the surface effects are absent, the dispersion branches for the symmetric SH waves will not cross the line $\Omega = \gamma$, while those for the antisymmetric SH waves still can. These phenomena are the same as those for piezoelectric materials. Furthermore, if we neglect the coupling between the electro-elastic field and the magnetic field, (48) degenerates to

$$\tanh\left(\frac{\pi}{2}\gamma\right) = \frac{\pi}{2}\alpha^2\gamma \quad (50)$$

which is the same as that in [47].

In the range $\gamma > \Omega$, we take the limit $\gamma \to \infty$, i.e., $\gamma$ is very large (the wavelength is very short compared to the bulk thickness of the plate), we have

$$\eta' \to \infty, \quad \tanh\left(\frac{\pi}{2}\eta'\right) \to 1, \quad \tanh\left(\frac{\pi}{2}\gamma\right) \to 1 \tag{51}$$

Therefore, both (46) and (47) become

$$\begin{vmatrix} \frac{\pi}{2}\left(a_1\gamma^2 - r_\rho\Omega^2\right)r_h + \eta' & \gamma\left(\frac{\pi}{2}a_2\gamma r_h - a_3\right) & \gamma\left(\frac{\pi}{2}a_4\gamma r_h - a_5\right) \\ a_3 + \frac{\pi}{2}e_{15}P_{12}^s\eta' r_h & a_6 + \frac{\pi}{2}a_7\gamma r_h & a_8 + \frac{\pi}{2}a_9\gamma r_h \\ a_5 + \frac{\pi}{2}h_{15}P_{13}^s\eta' r_h & a_8 + \frac{\pi}{2}a_{10}\gamma r_h & a_{11} + \frac{\pi}{2}a_{12}\gamma r_h \end{vmatrix} = 0, \tag{52}$$

which actually leads to the dispersion relation for SH surface waves in the magneto-electro-elastic nano-plate with surface effects. Discarding the surface effects, we obtain from (52)

$$\Omega^2 = \gamma^2\left(1 - l^2\right), \quad l = \frac{a_3\left(a_3 a_{11} - a_5 a_8\right) - a_5\left(a_3 a_8 - a_5 a_6\right)}{a_8^2 - a_6 a_{11}} \tag{53}$$

By further neglecting the coupling between the electro-elastic field and the magnetic field, the classical result of the Bleustein-Gulyaev wave [47, 49] can be recovered from (53). From (52) and (53), it is obvious that the surface wave in a magneto-electro-elastic plate without surface effects is non-dispersive, while it becomes dispersive when the surface effects are involved.

## 5. Numerical results and discussions

In order to quantitatively investigate the surface effects on the propagation characteristics of SH waves in magneto-electro-elastic nano-plates, numerical calculations have been conducted for a $BaTiO_3$-$CoFe_2O_4$ composite nano-plate. The material constants for the two constituents in the composite are given as follows [50]:

$BaTiO_3$:
$c_{44}^{(1)} = 4.3 \times 10^{10}$ Pa; $e_{15}^{(1)} = 11.6$ C/m$^2$; $\varepsilon_{11}^{(1)} = 1.12 \times 10^{-8}$ F/m;
$\mu_{11}^{(1)} = 5.0 \times 10^{-6}$ Ns$^2$/C$^2$; $\rho^{(1)} = 5.8 \times 10^3$ kg/m$^3$; $h_{15}^{(1)} = 0$ N/(A m)
$CoFe_2O_4$:
$c_{44}^{(2)} = 4.53 \times 10^{10}$ Pa; $h_{15}^{(2)} = 550$ N/(A m); $\varepsilon_{11}^{(2)} = 0.08 \times 10^{-9}$ F/m;
$\mu_{11}^{(2)} = -5.90 \times 10^{-4}$ Ns$^2$/C$^2$; $\rho^{(2)} = 5.3 \times 10^3$ kg/m$^3$; $e_{15}^{(2)} = 0$ C/m$^2$

The bulk material properties of the magneto-electro-elastic nano-plate are evaluated by the following rule of mixture:

$$M^b = f_e M_e^{(1)} + f_m M_m^{(2)} \tag{54}$$

where $M^b$ is any material constant of the bulk nano-plate, $M_e^{(1)}$ and $M_m^{(2)}$ are the corresponding material constants of BaTiO$_3$ and CoFe$_2$O$_4$, respectively; $f_e$ and $f_m$ which satisfy $f_e + f_m = 1$ are the volume fractions of BaTiO$_3$ and CoFe$_2$O$_4$ in the magneto-electro-elastic nano-plate. In the present work, we set $f_e = 0.8$ for the volume fraction of BaTiO$_3$. Note that the ME coupling is a unique property of the composite, since it is absent in both constituents. Therefore, in addition, the ME constant $\alpha_{11}$ for the bulk composite nano-plate is taken to be $\alpha_{11} = 0.005 \times 10^{-9}$ Ns/(VC) [50, 51]. The surface material parameters are evaluated from (33) when various material property ratios ($r_\rho$, $r_c$ and $r_e$) are specified.

For simplicity, we will use four abbreviations below, namely SMEE theory, 1$^{st}$ Anti-SHW, 1$^{st}$ Sym-SHW and SW, to denote the surface magneto-electro-elasticity theory, the lowest antisymmetric SH wave dispersion curve, the lowest symmetric SH wave dispersion curve, and the dispersion curve for the SH surface wave, respectively.

*5.1. Validation ranges of the SMEE theory*

To check the validation ranges of the proposed SMEE theory for SH waves, we first make a comparison of the approximate SMEE solution with the exact solution (EXACT) which is given in Appendix A. The accuracy of the SH wave dispersion relations determined by the present SMEE theory (36)-(39) and (46)-(47) is shown in figures 2 and 3 by the comparison with the exact SH wave dispersion relations which are governed by (A.11)-(A.14). For $r_\rho = 1$, $r_c = 0.5$ and $r_e = 5$, figures 2 and 3 compare the first 6 branches of the antisymmetric and symmetric SH wave dispersion curves, respectively, for four different values of $r_h$. It can be found from figures 2(a) and 3(a) that the SMEE theory agrees well with the exact solution for the 6 branches in the entire range of $\gamma < 8$ for $r_h = 0.05$. For $r_h = 0.1$, it can be seen from figures 2(b) and 3(b) that the SMEE theory is valid approximately for the second to fifth branches in the entire range $\gamma < 8$ and for the 1$^{st}$ Anti-SHW and 1$^{st}$ Sym-SHW in the range of $\gamma < 5$, respectively. However, the results are poor for higer branches and for the 1$^{st}$ Anti-SHW and 1$^{st}$ Sym-SHW in the range of $\gamma > 5$.

From figures 2(c) and 3(c), the SMEE theory for $r_h = 0.15$ is approximately effective for the second to fourth branches in the entire range of $\gamma < 8$ and for the 1$^{st}$ Anti-SHW and 1$^{st}$ Sym-SHW in the range of $\gamma < 3$, respectively. However, for higer branches and for the 1$^{st}$ Anti-SHW and 1$^{st}$ Sym-SHW in the range of $\gamma > 3$, the results are not satisfactory. Finally, from figures 2(d) and 3(d), it can be concluded that for $r_h = 0.2$ the SMEE theory gives a reasonable prediction only for the second and third branches in the entire range of $\gamma < 8$ and for the 1$^{st}$ Anti-SHW and 1$^{st}$ Sym-SHW in the range of $\gamma < 2$, respectively.

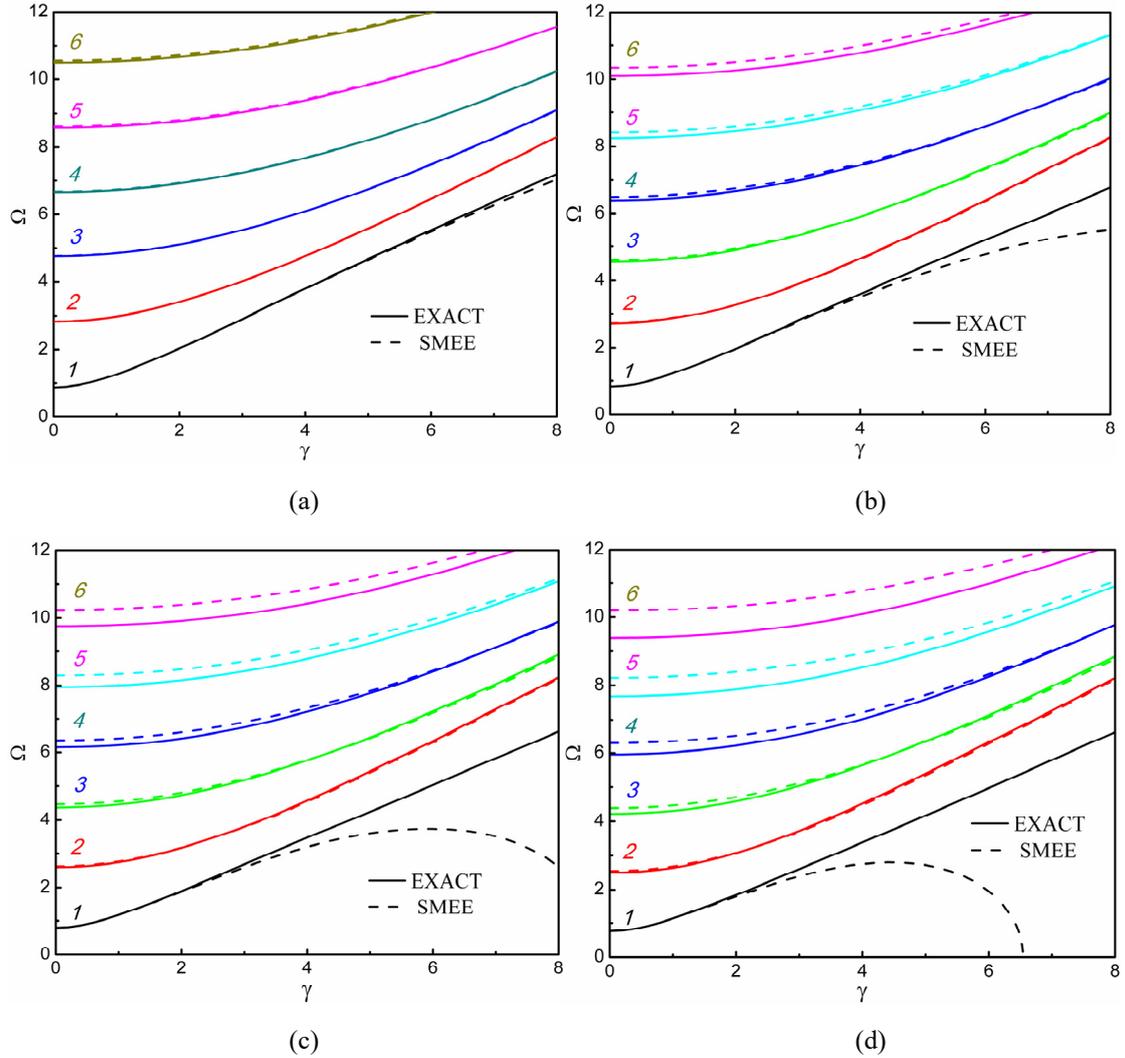

**Figure 2.** The comparison of the first 6 antisymmetric SH wave dispersion curves for different values of $r_h$ ($r_\rho = 1$, $r_c = 0.5$, $r_e = 5$): (a) $r_h = 0.05$; (b) $r_h = 0.1$; (c) $r_h = 0.15$; (d) $r_h = 0.2$.

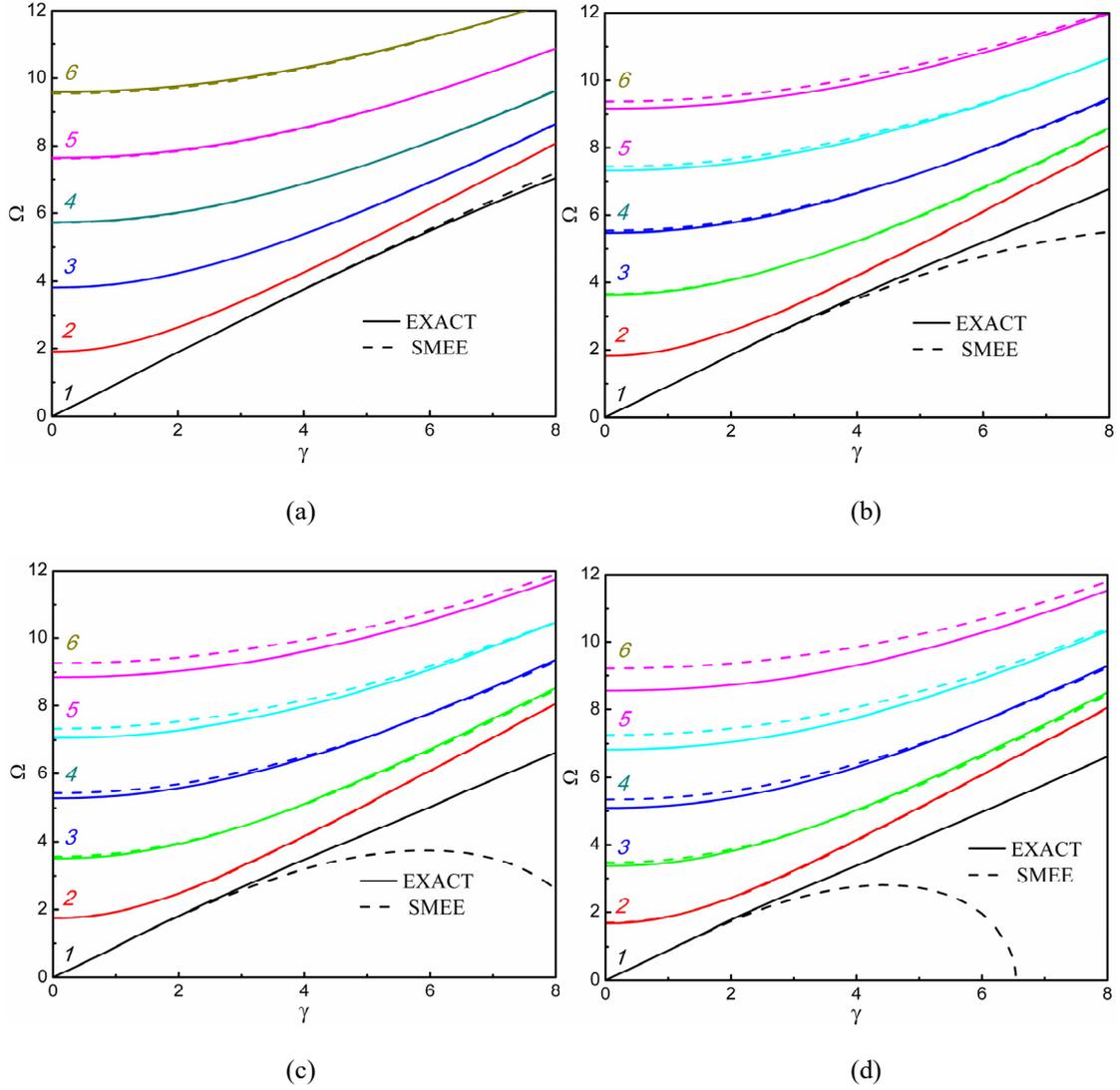

**Figure 3.** The comparison of the first 6 symmetric SH wave dispersion curves for different values of $r_h$ ($r_\rho = 1$, $r_c = 0.5$, $r_e = 5$): (a) $r_h = 0.05$; (b) $r_h = 0.1$; (c) $r_h = 0.15$; (d) $r_h = 0.2$.

In addition, for $r_\rho = 1$, $r_c = 0.5$ and $r_e = 5$, we also compare the results of the SH surface wave dispersion relations governed by the SMEE theory (52) with the exact SH surface wave dispersion relations (A.15) for four different values of $r_h$ in figure 4. It can be seen that, the validation ranges of the SH surface wave dispersion relations determined by the SMEE theory are similar to those of the 1st Anti-SHW and 1st Sym-SHW. Specifically, for $r_h = 0.05$, 0.1, 0.15 and 0.2, the SMEE theory agrees well with the exact solution for the SH surface wave in the range of $\gamma < 8$, 5, 3 and 2, respectively. In fact, if higher accuracy is required, then the higher-order SMEE theory should be employed by appropriately truncating the Taylor-series in (10).

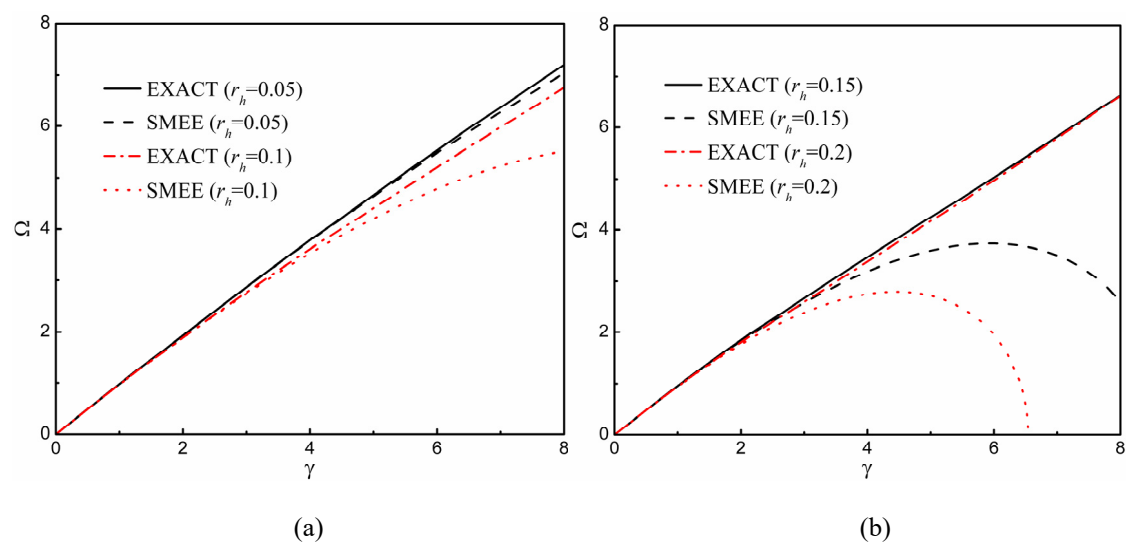

(a)                    (b)

**Figure 4.** The comparison of the SH surface wave dispersion curves for different values of $r_h$ ($r_\rho = 1$, $r_c = 0.5$, $r_e = 5$): (a) $r_h = 0.05$ and $r_h = 0.1$; (b) $r_h = 0.15$ and $r_h = 0.2$.

*5.2. Surface effects on SH wave propagation*

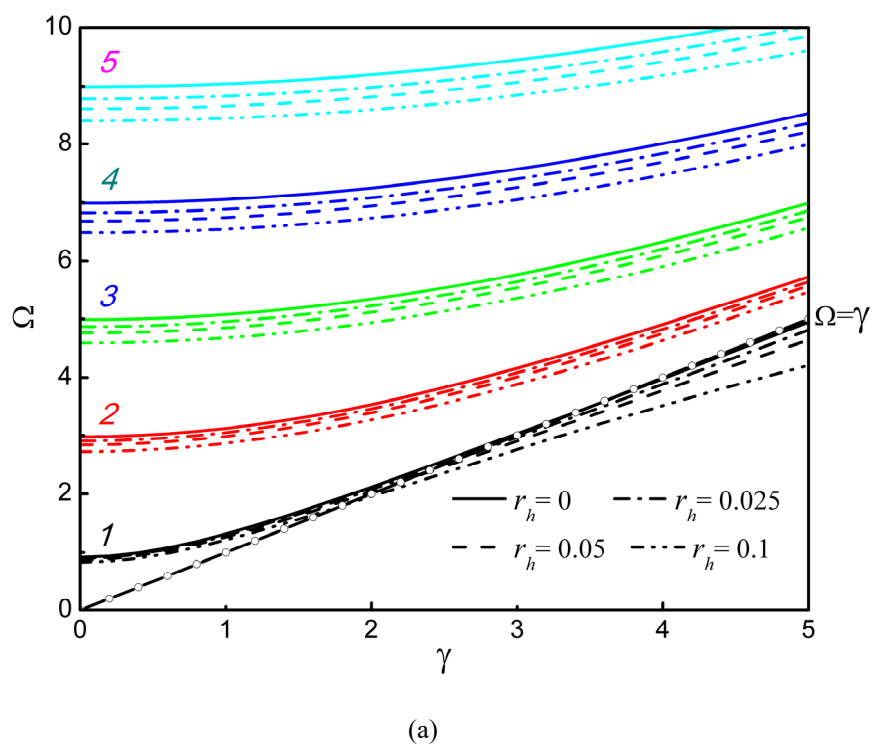

(a)

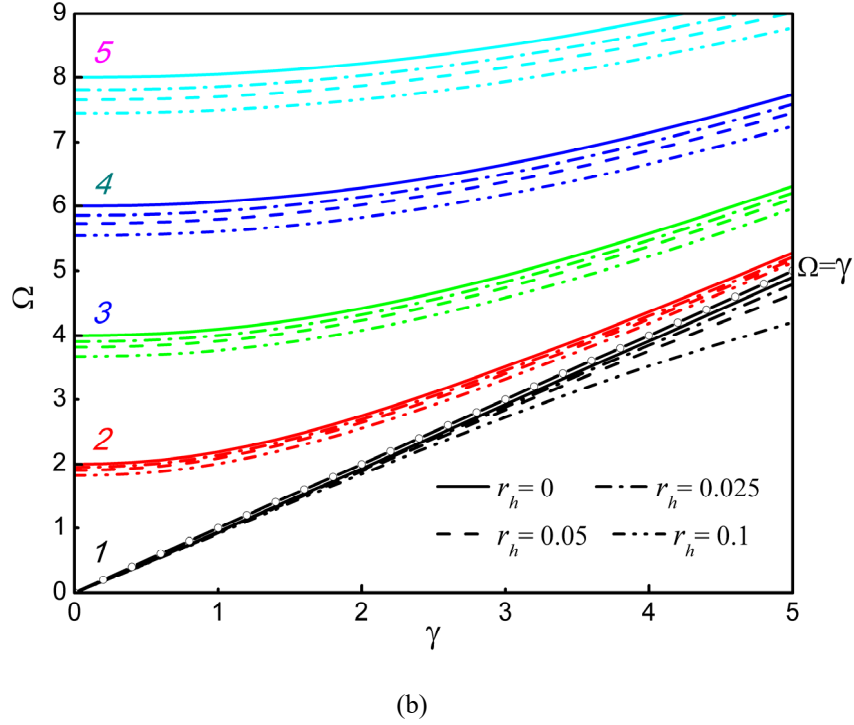

(b)

**Figure 5.** Dispersion curves of the first 5 SH wave modes for different values of $r_h$ ($r_\rho = 1$, $r_c = 0.5$, $r_e = 5$): (a) Antisymmetric; (b) Symmetric.

Below, the surface effects on SH waves propagating in magneto-electro-elastic nano-plates are numerically studied by the SMEE theory. As described above, the SMEE theory is valid approximately for the first 5 branches in the range of $\gamma < 5$ and $r_h < 0.1$. Therefore, for $r_\rho = 1$, $r_c = 0.5$ and $r_e = 5$, the first 5 branches of the antisymmetric and symmetric SH wave dispersion spectra for four different values of $r_h$ are depicted in figure 5. In particular, the results for $r_h = 0$ corresponding to no surface effects are included in figure 5 for comparison. It is found from figure 5 that for this combination of surface parameters, the frequencies for antisymmetric and symmetric SH waves are both reduced when surface effects are involved. With increasing $r_h$ (i.e., decreasing the thickness $H$ of the bulk material if the thickness $h$ of the surface layers is regarded as constant for a given nano-material), the dispersion curves deviate much more from those without surface effects. In addition, we can also observe that the surface effects on the first modes for both antisymmetric and symmetric SH waves are more significant at higher wave numbers than those at lower wave numbers. However, the surface effects on other high-order modes are quite different. The thickness-shear frequencies (corresponding to $\gamma = 0$) of high-order modes are significantly lowered.

From figure 5, it is also shown that for $r_\rho = 1$, $r_c = 0.5$ and $r_e = 5$, the lowest branch of the antisymmetric waves crosses the line $\Omega = \gamma$, while the lowest symmetric one does not. In order to clearly exhibit the relationships among the 1$^{st}$ Anti-SHW, 1$^{st}$ Sym-SHW, SW and the line $\Omega = \gamma$, we plot them for different values of $r_h$ in figure 6 with $r_\rho = 1$, $r_c = 0.5$ and $r_e = 5$.

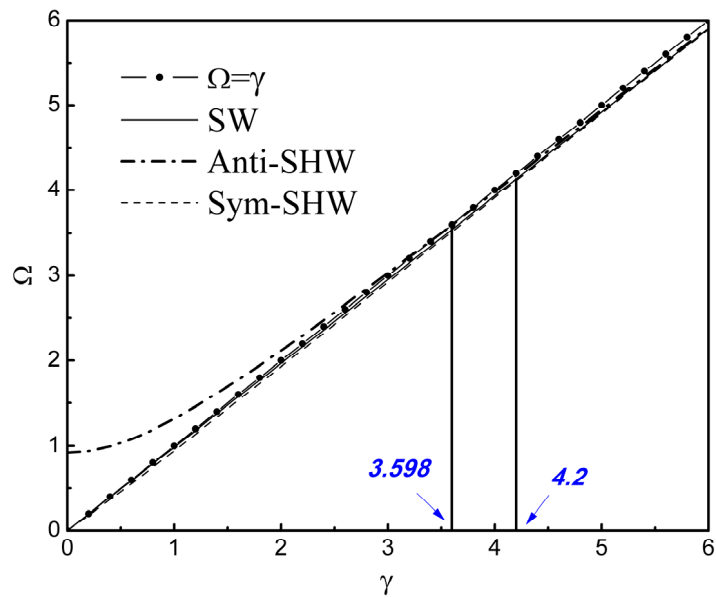

(a)

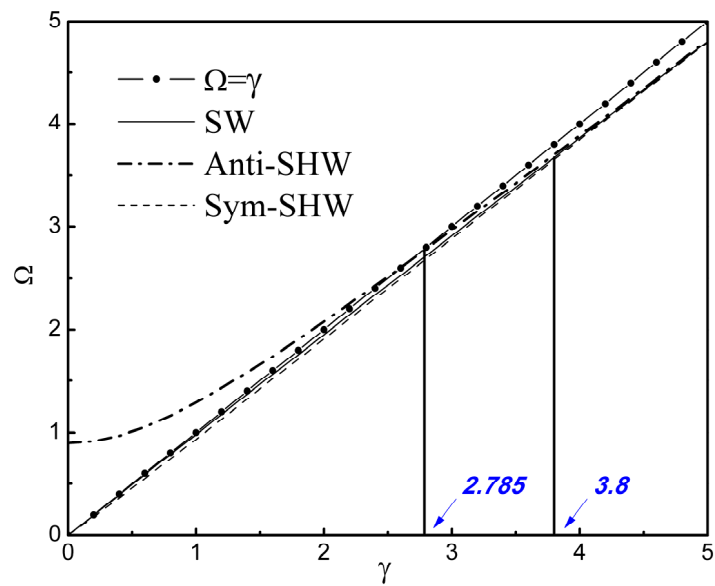

(b)

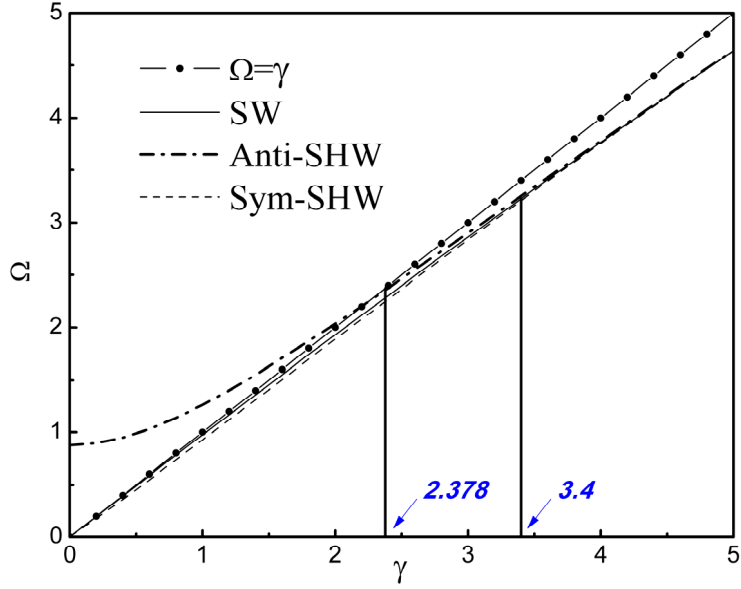

(c)

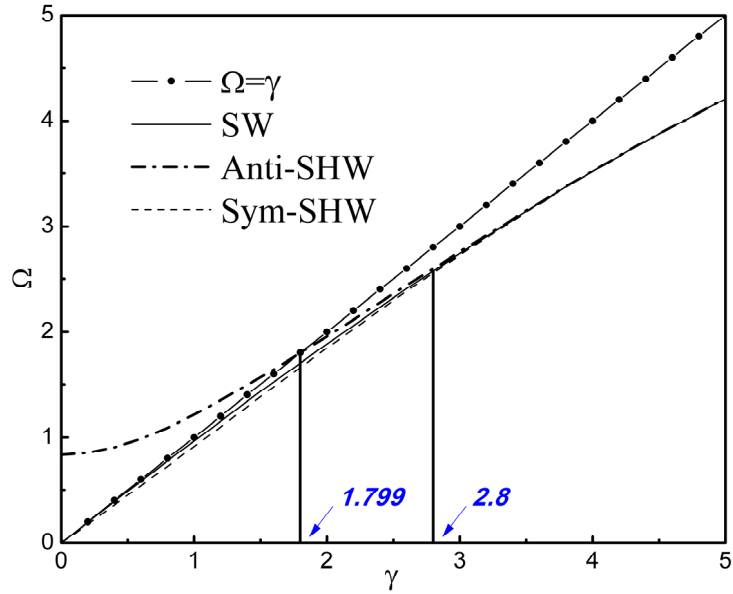

(d)

**Figure 6.** The 1$^{st}$ Anti-SHW, 1$^{st}$ Sym-SHW, SW and the line $\Omega = \gamma$ for different values of $r_h$ ($r_\rho = 1$, $r_c = 0.5$, $r_e = 5$): (a) $r_h = 0$; (b) $r_h = 0.025$; (c) $r_h = 0.05$; (d) $r_h = 0.1$.

For $r_h = 0$, 0.025, 0.05 and 0.1, it can be seen from figure 6 that the 1$^{st}$ Anti-SHW crosses the line $\Omega = \gamma$ at $\gamma = 3.598$, 2.785, 2.378 and 1.799, respectively, which are determined from (48). As shown in figure 6, the 1$^{st}$ Sym-SHW does not cross the line $\Omega = \gamma$ because there is no root of (49). Another interesting phenomenon is that the 1$^{st}$ Anti-SHW, 1$^{st}$ Sym-SHW and SW converge at certain wave number ($\gamma^* = 4.2$, 3.8, 3.4 and 2.8 for $r_h = 0$, 0.025, 0.05 and 0.1, respectively), which means that after these wave numbers, both the 1st Anti-SHW and 1st Sym-SHW have the character of a surface wave, i.e., the amplitude of the displacement is greatest near the surface of the nano-plate and decays exponentially into the interior [47].

To investigate the surface effects on the SH surface wave propagation, for $r_\rho = 1$, $r_c = 0.5$ and $r_e = 5$, the SW for four different values of $r_h$ is shown in figure 7, where the dimensionless phase velocity $V$ is defined as $V = \Omega/\gamma = c/c_T$, with $c = \omega/k$ and $c_T = \sqrt{c_{44}^*/\rho}$ denoting the phase velocity and the transverse wave velocity in the bulk material.

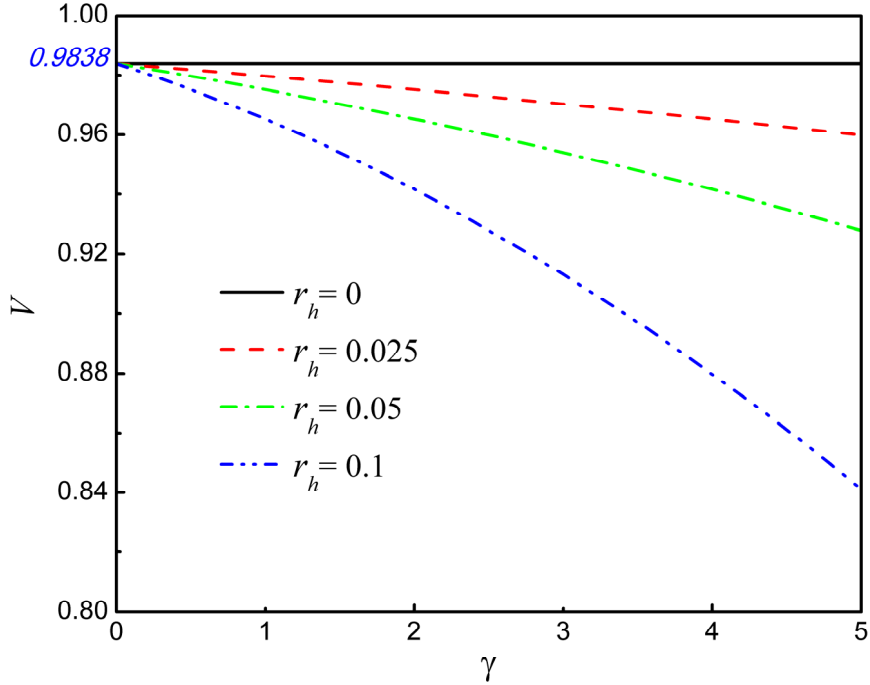

**Figure 7.** The SW for different values of $r_h$ ($r_\rho = 1$, $r_c = 0.5$, $r_e = 5$).

It is obvious from figure 7 that the SH surface wave in a magneto-electro-elastic plate without surface effects (corresponding to $r_h = 0$) propagates at a constant phase velocity $c = 0.9838$ determined by (53), and hence is non-dispersive. However, it becomes dispersive when the surface effects are involved. In addition, the dispersion curves deviate much more from those without surface effects with increasing $r_h$. We can also observe that the surface effects on the SH surface wave are more significant at higher wave numbers than those at lower wave numbers which is similar to the 1st Anti-SHW and 1st Sym-SHW in figure 5.

In order to clearly display the surface effects on the SH wave propagation behavior, the curves of the dimensionless frequency of the second symmetric SH wave mode (for $\gamma = 0$ and $\gamma = 4$, respectively) versus $\log_{10}(H/h)$ are shown in figures 8(a) and 8(b) for different values of $r_\rho$. As shown in figure 3, the validation ranges of the SMEE theory for the second symmetric branch is $r_h < 0.2$, i.e., $\log_{10}(H/h) > 0.699$, in the entire range of $\gamma < 8$. As a consequence, we take the minimum value of $\log_{10}(H/h)$ as 0.7. It is worth mentioning that, the thickness-shear mode frequencies for the symmetric SH waves at $\gamma = 0$ are independent of $r_c$ and $r_e$ as seen from (41). Therefore, only $r_\rho$ is varied to modulate the surface property as shown in figure 8(a). However, for $\gamma = 4$, the dimensionless frequencies further depend on $r_c$ and $r_e$. Thus, we have specified $r_c = 2$ and $r_e = 5$ to obtain the results in figure 8(b).

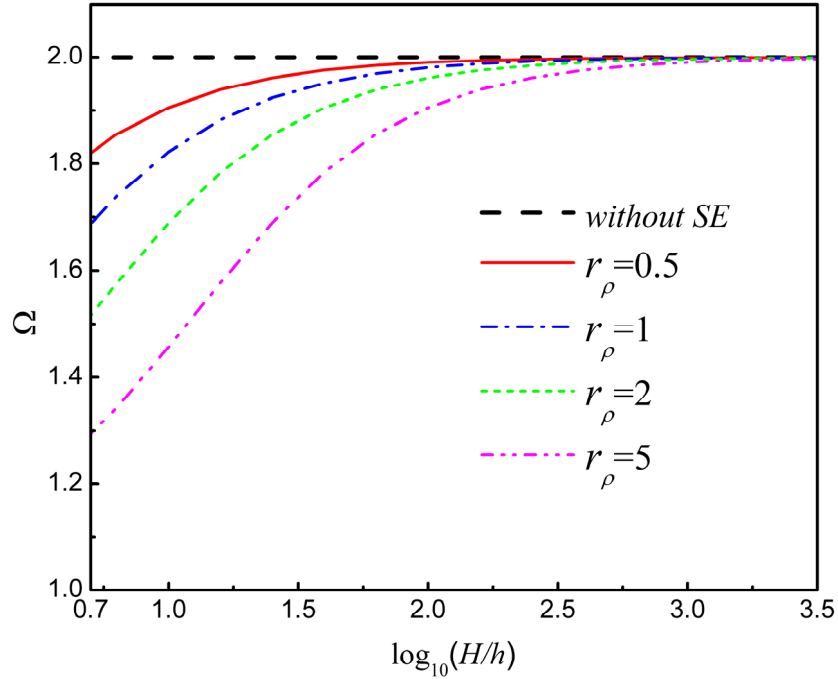

(a)

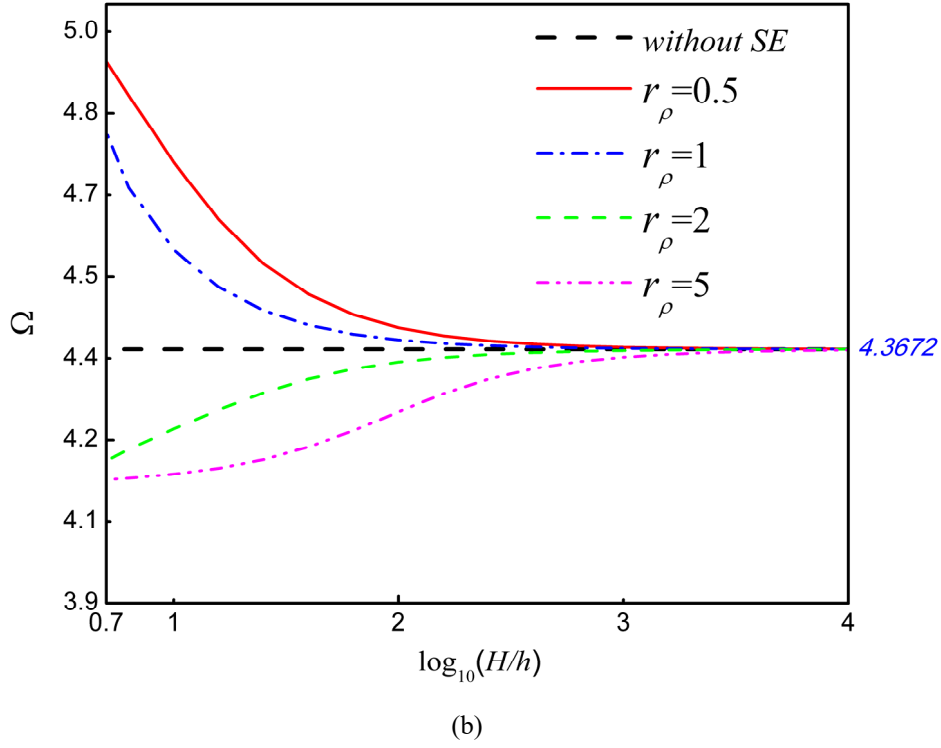

(b)

**Figure 8.** Dimensionless frequency of the second symmetric SH wave mode versus $\log_{10}(H/h)$ for different values of $r_\rho$: (a) $\gamma = 0$; (b) $\gamma = 4$, $r_c = 2$, $r_e = 5$.

It can be seen from figure 8 that the frequencies for $\gamma = 0$ and $\gamma = 4$ approach those without surface effects (i.e., $\Omega = 2$ and 4.3672, respectively) when $H/h$ becomes infinitely large (i.e., the bulk layer thickness of the plate $H \to \infty$), which is expected since the surface effects become trivial at macroscale. As a result, for the macroscopic magneto-electro-elastic plate, the surface effects may be neglected. However, it can be found that the frequencies begin to deviate from those without surface effects at $r_h = 0.003$ and 0.001 corresponding to figures 8(a) and 8(b), respectively. Thus, when the bulk layer thickness decreases to nanoscale, the surface effects may be significant and should be considered in the modeling [35, 37]. Both Figures 8(a) and 8(b) also indicate that for a fixed $H/h$, the surface effects lower the thickness-shear frequencies of the second symmetric SH wave mode as $r_\rho$ increases. However, when surface effects are involved, the thickness-shear frequencies for $0.5 < r_\rho < 5$ are always less than those without surface effects for figure 8(a). On the contrary, for figure 8(b), the frequencies are not always less than those without surface effects. Specifically, the frequencies for $r_\rho = 0.5$ and 1 are more than those without surface effects while the frequencies for $r_\rho = 2$ and 5 are less than. Therefore, by varying the values of $r_\rho$ to change the surface property, we can modulate the SH wave propagation in nano-plates.

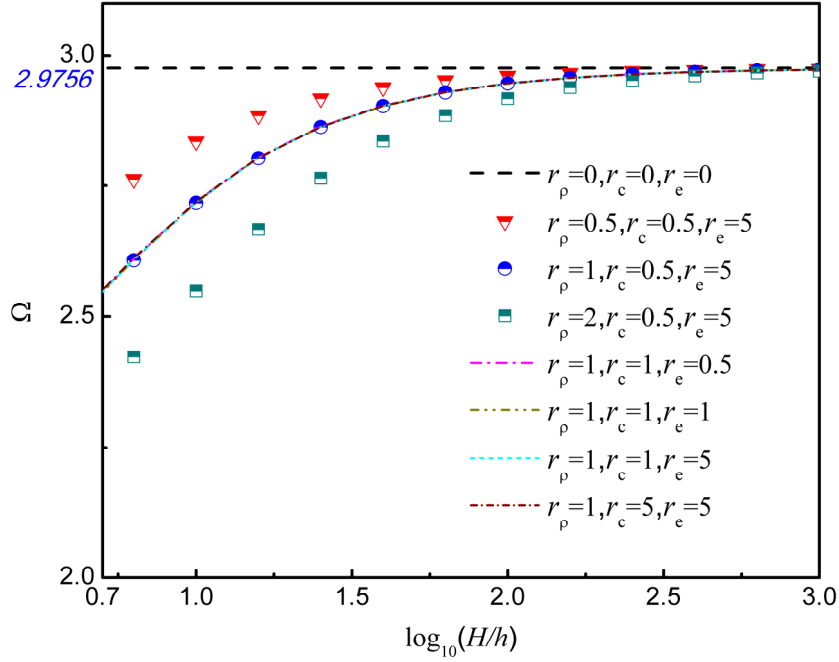

**Figure 9.** Dimensionless thickness-shear frequency of the second symmetric SH wave mode versus $\log_{10}(H/h)$ for different surface parameters.

The surface effects on the thickness-shear frequencies of the antisymmetric SH wave modes are more complicated, because they depend on $r_\rho$, $r_c$ and $r_e$ according to (40). The dimensionless thickness-shear frequency of the second antisymmetric SH wave mode versus $\log_{10}(H/h)$ is plotted in figure 9 for different combinations of $r_\rho$, $r_c$ and $r_e$. We also take the minimum value of $\log_{10}(H/h)$ as 0.7. It can be seen from figure 9 that, when $H/h$ becomes infinitely large, the thickness-shear frequency also approaches that without surface effects (i.e., $\Omega = 2.9756$). When $r_\rho$ varies, figure 9 shows similar phenomena to those in figure 8(a). In addition, it can be seen from figure 8 that if the surface effects are involved, the curve of thickness-shear frequency versus $\log_{10}(H/h)$ for the second antisymmetric SH wave is insensitive to $r_c$ and $r_e$. That is to say, the thickness-shear frequency of the second antisymmetric SH wave mainly depends on $r_\rho$ once $r_h$ is specified. For the higher wave numbers of the second antisymmetric SH wave, we can obtain similar results to those in figures 8(b) and 9.

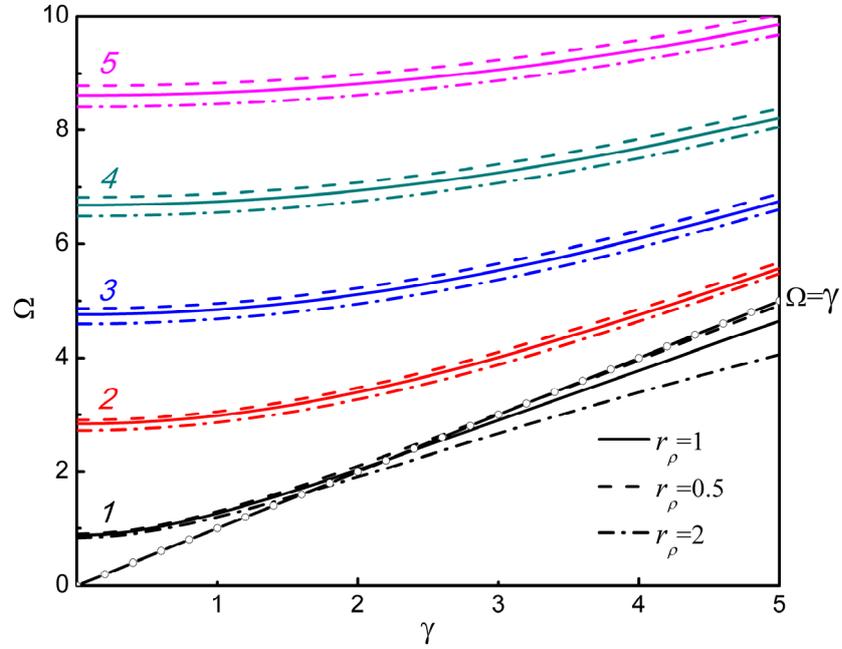

(a)

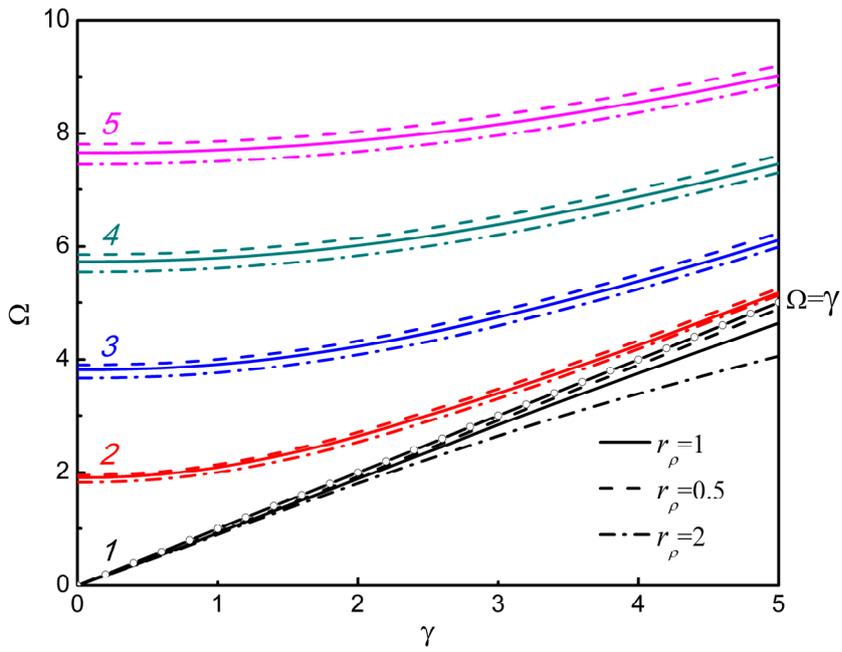

(b)

**Figure 10.** Dispersion curves of the first 5 SH wave modes for different values of $r_\rho$ ($r_h = 0.05$, $r_c = 0.5$, $r_e = 5$): (a) Antisymmetric; (b) Symmetric.

For $r_h = 0.05$, $r_c = 0.5$ and $r_e = 5$, the first 5 branches of the antisymmetric and symmetric SH wave dispersion spectra for different values of $r_\rho$ are depicted in figure 10. It is noted from figure 10 that increasing the density ratio $r_\rho$ will decrease the frequencies of both antisymmetric and symmetric SH wave modes. That is to say, the greater is the surface density, the lower is the frequency. In addition, we can also observe that the surface effects on the first modes for both antisymmetric and symmetric SH waves are more significant at higher wave numbers than those at lower wave numbers, which is similar to figure 5.

For $r_h = 0.05$, $r_\rho = 2$ and $r_c = 0.5$, the first 5 branches of the antisymmetric and symmetric SH wave dispersion spectra for different values of $r_e$ are depicted in figure 11. It can be seen that, when the surface effects are taken into account, the variation of $r_e$ gives only little impact on the dispersion curves for both antisymmetric and symmetric SH wave modes. That is to say, both antisymmetric and symmetric SH wave is insensitive to $r_e$ from figure 11, which shows similar phenomena to those in figure 9.

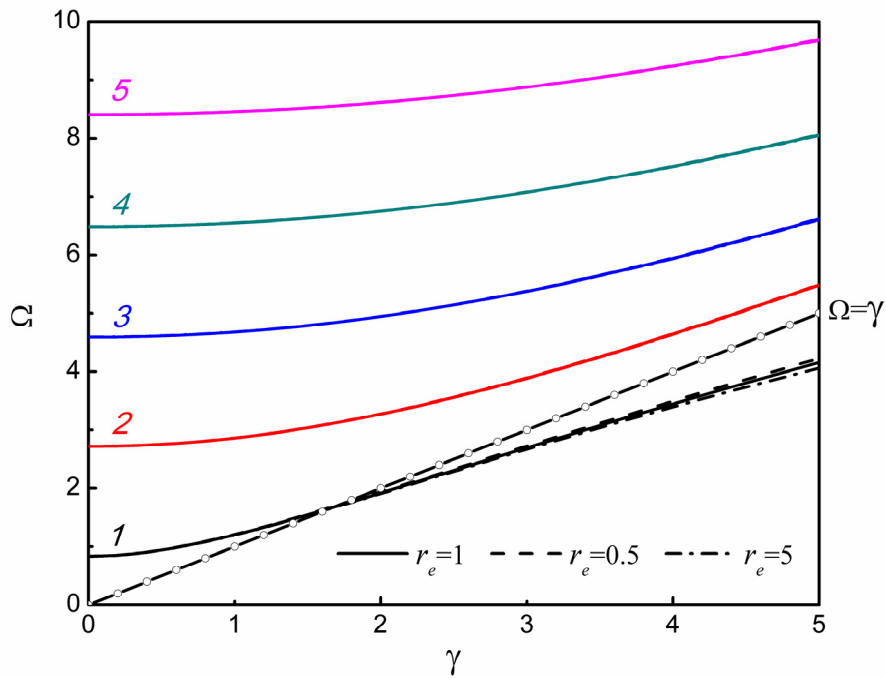

(a)

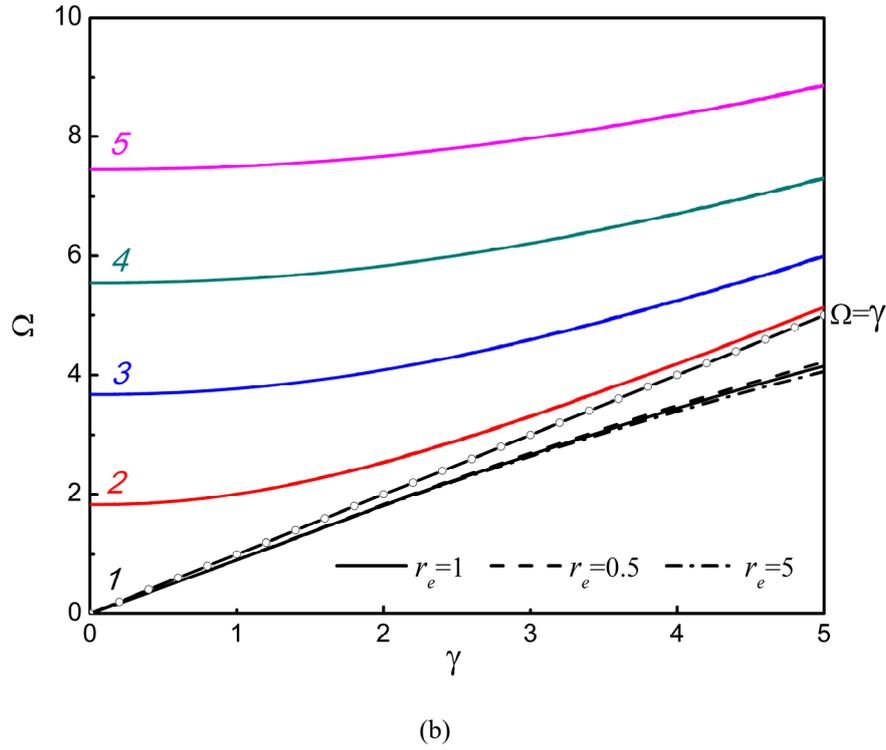

(b)

**Figure 11.** Dispersion curves of the first 5 SH wave modes for different values of $r_e$ ($r_h = 0.05$, $r_\rho = 2$, $r_c = 0.5$): (a) Antisymmetric; (b) Symmetric.

For $r_h = 0.05$, $r_\rho = 2$ and $r_e = 5$, the first 5 branches of the antisymmetric and symmetric SH wave dispersion spectra for different values of $r_c$ are shown in figure 12. It is found that, for high-order modes, a soft surface corresponding to $r_c = 0.5$ has only little influence on the dispersion curves in the entire wave number range for higher-order antisymmetric and symmetric SH waves. On the contrary, although the dispersion curves are also not affected by a stiff surface corresponding to $r_c = 5$ at relatively low wave numbers for high-order modes, it plays a significant role at higher wave numbers. Both the 1st Anti-SHW and the 1st Sym-SHW are very sensitive to $r_c$, especially at high wave numbers. Increasing the stiffness of the surface (i.e., increasing $r_c$) will obviously increase the frequencies of both the 1st Anti-SHW and the 1st Sym-SHW.

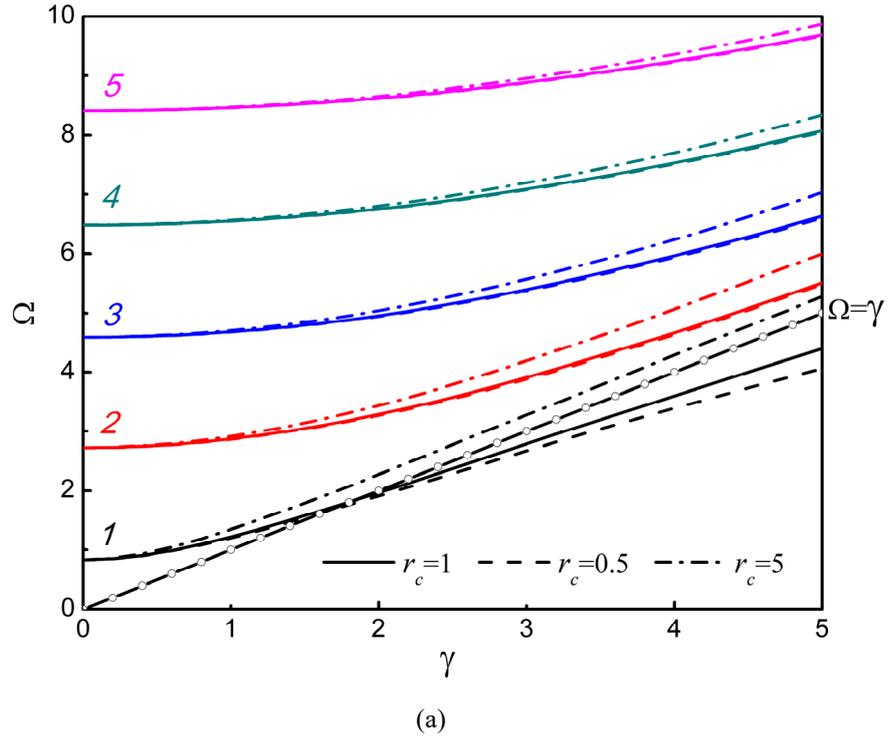

(a)

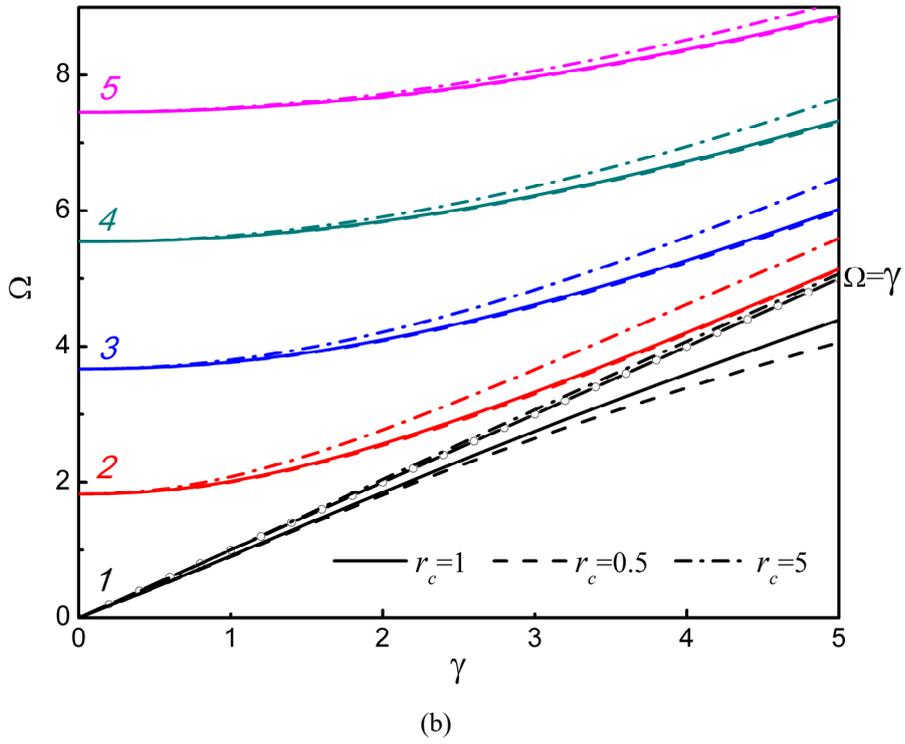

(b)

**Figure 12.** Dispersion curves of the first 5 SH wave modes for different values of $r_c$ ($r_h = 0.05$, $r_\rho = 2$, $r_e = 5$): (a) Antisymmetric; (b) Symmetric.

Another interest phenomenon is displayed in figure 13 which clearly shows the relationships among the 1st Anti-SHW, 1st Sym-SHW and the line $\Omega = \gamma$ for $r_h = 0.05$, $r_\rho = 2$, $r_c = 5$ and $r_e = 5$. It can be seen from figure 13 that the 1st Sym-SHW will cross the line $\Omega = \gamma$ at $\gamma = 0.97$ which is determined from (49), while the 1st Anti-SHW will not cross the line $\Omega = \gamma$ because there is no root of (48). This phenomenon is solely induced by the surface effects and is different from the case without surface effects. Thus, by changing the stiffness of the surface, we also can modulate the SH wave propagation.

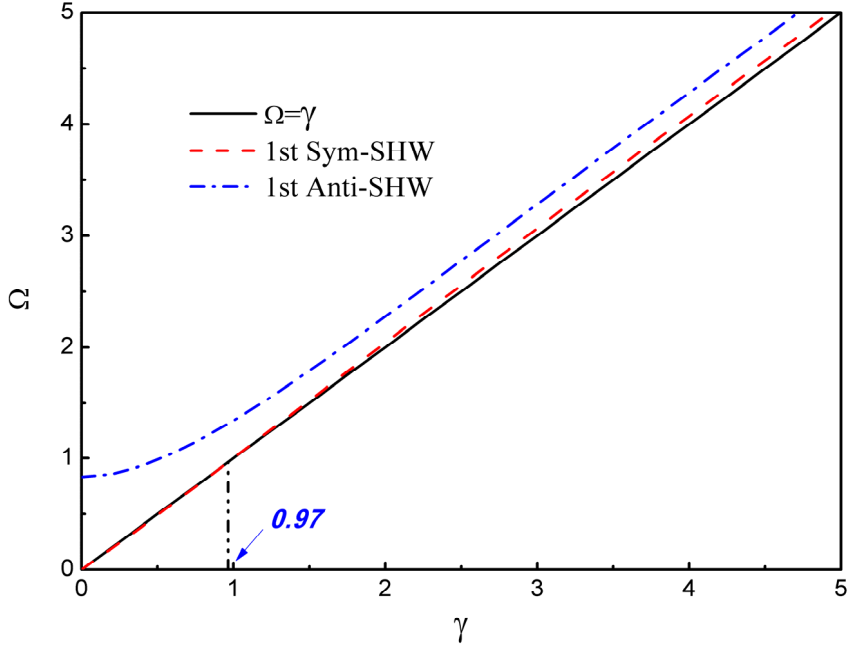

**Figure 13.** The 1st Anti-SHW, 1st Sym-SHW and the line $\Omega = \gamma$ ($r_h = 0.05$, $r_\rho = 2$, $r_c = 5$, $r_e = 5$).

## 6. Conclusions

In this study, surface effects on anti-plane shear (SH) waves propagating in a transversely isotropic magneto-electro-elastic nano-plate are analyzed. First, the state-space formalism for the anti-plane problem is used to develop a surface magneto-electro-elasticity theory for magneto-electro-elastic surface layers under traction-free, magnetically open and electrically shorted boundary conditions, in which the thickness of the surface layer is involved. Based on the developed multi-field coupled surface theory, the size-dependent dispersion relations for anti-symmetric and symmetric SH waves are derived analytically. In particular, the relations for the special thickness-shear modes as well as the asymptotic characteristics of the dispersion relations with surface effects are obtained. Finally, numerical examples for a $BaTiO_3$-$CoFe_2O_4$ composite nano-plate are presented and discussed to determine the validation ranges of the SMEE theory and highlight the surface effects on the dispersion spectra of SH waves.

The behavior of the SH waves with surface effects is complex and depends strongly on the considered mode, the considered frequency range, and the chosen surface material parameters. Specifically, the surface effects on the $1^{st}$ Anti-SHW, $1^{st}$ Sym-SHW and SW are more significant at higher wave numbers than those at lower wave numbers. Furthermore, we have shown that the $1^{st}$ Anti-SHW and $1^{st}$ Sym-SHW converge at certain wave number after which these two modes both show the character of a surface SH wave. In addition, when the bulk layer thickness of the plate decreases to nanoscale, the frequencies deviate promptly from those without surface effects. Besides, the greater is the surface density, the lower is the frequency. Both antisymmetric and symmetric SH waves are insensitive to the surface ME coupling. The soft surface layer has only a small effect on the dispersion relations for higher-order antisymmetric and symmetric modes in the considered wave number range, while the $1^{st}$ Anti-SHW and $1^{st}$ Sym-SHW are more senstive to them, especially at a higher wave number. For high-order modes, the difference between the dispersion relations with and without surface effects is more significant for stiff surface layers than for soft surface layers. In summary, the size-dependent dispersion relations demonstrate that the surface effects are very significant and should be considered for accurately modeling the physical properties of magneto-electro-elastic nano-plates. As a consequence, it is possible to modulate the elastic waves in magneto-electro-elastic nano-plates through surface engineering. This work provides a theoretical guidance for the design and applications of multiferroic nanoplate-based devices in Nano-Electro-Mechanical Systems (NEMS).

It is emphasized here that if higher accuracy is required, then the higher-order SMEE theory should be employed. Furthermore, the SMME theory corresponding to different boundary conditions and different surface configurations can be obtained in a similar way. In addition, the proposed method can be conveniently applied in the development of accurate interface magneto-electro-elastic theories.

Finally, it should be remarked that the model described in Section 2 as shown in Fig. 1 is not necessary a nano-plate, and it could be a general plate with a core (bulk material) and two cover layers in the macro-scale range. However, the main objective of this paper is to reveal the surface effects on the wave propagation characteristics from the physical points of view, which are important in nano-structures. If the plate is considered as a magneto-electro-elastic composite structure at macro-scales, then the surface effects can be neglected and we do not need to consider the two surface layers and only need to take the individual bulk material into account. On the other hand, when the plate is considered as a macro-scaled core material covered by two surface coatings whose properties are different from the bulk material, then the present model can be used to analyze the effects of the surface coatings.

**Acknowledgments**

The work was supported by the National Natural Science Foundation of China (Nos. 11090333, 11272281, 11321202 and 11202182), the German Research Foundation (DFG, Project-No: ZH 15/20-1), and the China Scholarship Council (CSC).

**Appendix A: Exact SH wave dispersion relations in three-layer magneto-electro-elastic plates**

If we treat the magneto-electro-elastic nano-plate as a three-layer composite structure with both the two identical surface layers and the bulk material modeled directly by the exact three-dimensional magneto-electro-elasticity theory, then the exact dispersion relations can be obtained for the SH waves propagating in the three-layer composite structure. The wave solution in the bulk material of the nano-plate is represented by (22)-(25). However, the wave solutions in the top and bottom surface layers can be expressed as the following forms

$$u_3^{(j)} = A_j \sin(\beta^s x_2) + B_j \cos(\beta^s x_2), \quad \xi^{(j)} = C_j \sinh(kx_2) + D_j \cosh(kx_2),$$
$$\zeta^{(j)} = E_j \sinh(kx_2) + F_j \cosh(kx_2), \quad (\beta^s)^2 = (\rho^s \omega^2 - c_{44}^{*s} k^2)/c_{44}^{*s}, \quad \text{(A.1)}$$

$$\phi^{(j)} = m_1^s \left[ A_j \sin(\beta^s x_2) + B_j \cos(\beta^s x_2) \right] + b_1^s \left[ C_j \sinh(kx_2) + D_j \cosh(kx_2) \right]$$
$$+ b_2^s \left[ E_j \sinh(kx_2) + F_j \cosh(kx_2) \right],$$
$$\psi^{(j)} = m_2^s \left[ A_j \sin(\beta^s x_2) + B_j \cos(\beta^s x_2) \right] + b_2^s \left[ C_j \sinh(kx_2) + D_j \cosh(kx_2) \right] \quad \text{(A.2)}$$
$$+ b_3^s \left[ E_j \sinh(kx_2) + F_j \cosh(kx_2) \right],$$

$$T_{23}^{(j)} = c_{44}^{*s} \beta^s \left[ A_j \cos(\beta^s x_2) - B_j \sin(\beta^s x_2) \right] - m_1^s k \left[ C_j \cosh(kx_2) + D_j \sinh(kx_2) \right]$$
$$- m_2^s k \left[ E_j \cosh(kx_2) + F_j \sinh(kx_2) \right], \quad \text{(A.3)}$$
$$D_2^{(j)} = k \left[ C_j \cosh(kx_2) + D_j \sinh(kx_2) \right], \quad B_2^{(j)} = k \left[ E_j \cosh(kx_2) + F_j \sinh(kx_2) \right]$$

where the superscript $j$ takes 1 and 2 representing the wave solutions of the top and the bottom surface layers, respectively. Here again, the common factor $\exp[i(kx_1 - \omega t)]$ is omitted in all field variables.

The traction-free, magnetically open and electrically shorted boundary conditions for the top and bottom surface layers at $x_2 = H + h$ and $x_2 = -H - h$ can be written as

$$\mathbf{T}^{(1)}(H+h) = 0, \quad \mathbf{T}^{(2)}(-H-h) = 0 \quad \text{(A.4)}$$

The state variables of the top and bottom surface layers should be equal to those of the bulk material at $x_2 = H$ and $x_2 = -H$, i.e.,

$$\mathbf{T}^{(1)}(H) = \mathbf{T}(H), \quad \mathbf{u}^{(1)}(H) = \mathbf{u}(H), \quad \mathbf{T}^{(2)}(-H) = \mathbf{T}(-H), \quad \mathbf{u}^{(2)}(-H) = \mathbf{u}(-H) \quad \text{(A.5)}$$

Substituting (22)$_1$, (24), (25) and (A.1)-(A.3) into the boundary conditions (A.4) and the continuity conditions (A.5), adding and subtracting the resulting equations, then two sets of nine homogeneous equations can be obtained as follows

$$\mathbf{G}^{anti} \left[ A, C, E, A_1 + A_2, B_1 - B_2, C_1 + C_2, D_1 - D_2, E_1 + E_2, F_1 - F_2 \right]^T = \mathbf{0},$$
$$\mathbf{G}^{sym} \left[ B, D, F, A_1 - A_2, B_1 + B_2, C_1 - C_2, D_1 + D_2, E_1 - E_2, F_1 + F_2 \right]^T = \mathbf{0} \quad \text{(A.6)}$$

where $G_{ij}^{anti}$ and $G_{ij}^{sym}$ are the elements of the matrices $\mathbf{G}^{anti}$ and $\mathbf{G}^{sym}$, respectively,

which are given by

$$[G_{ij}^{anti}] = \begin{bmatrix} 0 & 0 & 0 & c_{44}^{*s}\beta^s c_3 & -c_{44}^{*s}\beta^s s_3 & -m_1^s k c_5 & -m_1^s k s_5 & -m_2^s k c_5 & -m_2^s k s_5 \\ 0 & 0 & 0 & m_1^s s_3 & m_1^s c_3 & b_1^s s_5 & b_1^s c_5 & b_2^s s_5 & b_2^s c_5 \\ 0 & 0 & 0 & m_2^s s_3 & m_2^s c_3 & b_2^s s_5 & b_2^s c_5 & b_3^s s_5 & b_3^s c_5 \\ s_1 & 0 & 0 & s_2 & c_2 & 0 & 0 & 0 & 0 \\ m_1 s_1 & b_1 s_4 & b_2 s_4 & m_1^s s_2 & m_1^s c_2 & b_1^s s_4 & b_1^s c_4 & b_2^s s_4 & b_2^s c_4 \\ m_2 s_1 & b_2 s_4 & b_3 s_4 & m_2^s s_2 & m_2^s c_2 & b_2^s s_4 & b_2^s c_4 & b_3^s s_4 & b_3^s c_4 \\ c_{44}^* \beta c_1 & -m_1 k c_4 & -m_2 k c_4 & c_{44}^{*s}\beta^s c_2 & -c_{44}^{*s}\beta^s s_2 & -m_1^s k c_4 & -m_1^s k s_4 & -m_2^s k c_4 & -m_2^s k s_4 \\ 0 & c_4 & 0 & 0 & 0 & c_4 & s_4 & 0 & 0 \\ 0 & 0 & c_4 & 0 & 0 & 0 & 0 & c_4 & s_4 \end{bmatrix} \quad (A.7)$$

and

$$[G_{ij}^{sym}] = \begin{bmatrix} 0 & 0 & 0 & c_{44}^{*s}\beta^s c_3 & -c_{44}^{*s}\beta^s s_3 & -m_1^s k c_5 & -m_1^s k s_5 & -m_2^s k c_5 & -m_2^s k s_5 \\ 0 & 0 & 0 & m_1^s s_3 & m_1^s c_3 & b_1^s s_5 & b_1^s c_5 & b_2^s s_5 & b_2^s c_5 \\ 0 & 0 & 0 & m_2^s s_3 & m_2^s c_3 & b_2^s s_5 & b_2^s c_5 & b_3^s s_5 & b_3^s c_5 \\ c_1 & 0 & 0 & s_2 & c_2 & 0 & 0 & 0 & 0 \\ m_1 c_1 & b_1 c_4 & b_2 c_4 & m_1^s s_2 & m_1^s c_2 & b_1^s s_4 & b_1^s c_4 & b_2^s s_4 & b_2^s c_4 \\ m_2 c_1 & b_2 c_4 & b_3 c_4 & m_2^s s_2 & m_2^s c_2 & b_2^s s_4 & b_2^s c_4 & b_3^s s_4 & b_3^s c_4 \\ c_{44}^* \beta s_1 & m_1 k s_4 & m_2 k s_4 & -c_{44}^{*s}\beta^s c_2 & c_{44}^{*s}\beta^s s_2 & m_1^s k c_4 & m_1^s k s_4 & m_2^s k c_4 & m_2^s k s_4 \\ 0 & s_4 & 0 & 0 & 0 & c_4 & s_4 & 0 & 0 \\ 0 & 0 & s_4 & 0 & 0 & 0 & 0 & c_4 & s_4 \end{bmatrix} \quad (A.8)$$

where

$$\begin{aligned} & s_1 = \sin(\beta H), \quad s_2 = \sin(\beta^s H), \quad s_3 = \sin[\beta^s(H+h)], \quad s_4 = \sinh(kH), \\ & s_5 = \sinh[k(H+h)], \quad c_1 = \cos(\beta H), \quad c_2 = \cos(\beta^s H), \\ & c_3 = \cos[\beta^s(H+h)], \quad c_4 = \cosh(kH), \quad c_5 = \cosh[k(H+h)] \end{aligned} \quad (A.9)$$

For nontrivial solutions, the determinant of the coefficient matrices in (A.6) should vanish, giving rise to the following frequency equations

$$|\mathbf{G}^{anti}| = 0, \quad |\mathbf{G}^{sym}| = 0, \quad (A.10)$$

which determine the exact dispersion relations for the antisymmetric and symmetric SH waves in the three-layer magneto-electro-elastic nano-plate, respectively.

Substituting (32) and (33)$_{1,2}$ into (A.7)-(A.10) and after some rearrangements, we can obtain the dimensionless forms of the exact dispersion relations (A.10) as

$$|\overline{\mathbf{G}}^{anti}| = 0, \quad |\overline{\mathbf{G}}^{sym}| = 0 \quad (A.11)$$

where

$$\left[\bar{G}_{ij}^{anti}\right] = \begin{bmatrix} 0 & 0 & 0 & \eta^s\bar{c}_3 & -\eta^s\bar{s}_3 & -g_2\gamma\bar{c}_5 & -g_2\gamma\bar{s}_5 & -g_3\gamma\bar{c}_5 & -g_3\gamma\bar{s}_5 \\ 0 & 0 & 0 & g_1\bar{s}_3 & g_1\bar{c}_3 & g_5\bar{s}_5 & g_5\bar{c}_5 & g_6\bar{s}_5 & g_6\bar{c}_5 \\ 0 & 0 & 0 & g_4\bar{s}_3 & g_4\bar{c}_3 & g_6\bar{s}_5 & g_6\bar{c}_5 & g_7\bar{s}_5 & g_7\bar{c}_5 \\ \bar{s}_1 & 0 & 0 & \bar{s}_2 & \bar{c}_2 & 0 & 0 & 0 & 0 \\ a_3\bar{s}_1 & a_6\bar{s}_4 & a_8\bar{s}_4 & g_1\bar{s}_2 & g_1\bar{c}_2 & g_5\bar{s}_4 & g_5\bar{c}_4 & g_6\bar{s}_4 & g_6\bar{c}_4 \\ a_5\bar{s}_1 & a_8\bar{s}_4 & a_{11}\bar{s}_4 & g_4\bar{s}_2 & g_4\bar{c}_2 & g_6\bar{s}_4 & g_6\bar{c}_4 & g_7\bar{s}_4 & g_7\bar{c}_4 \\ \eta\bar{c}_1 & -a_3\gamma\bar{c}_4 & -a_5\gamma\bar{c}_4 & g_8\eta^s\bar{c}_2 & -g_8\eta^s\bar{s}_2 & -g_1\gamma\bar{c}_4 & -g_1\gamma\bar{s}_4 & -g_4\gamma\bar{c}_4 & -g_4\gamma\bar{s}_4 \\ 0 & \bar{c}_4 & 0 & 0 & 0 & \bar{c}_4 & \bar{s}_4 & 0 & 0 \\ 0 & 0 & \bar{c}_4 & 0 & 0 & 0 & 0 & \bar{c}_4 & \bar{s}_4 \end{bmatrix} \quad (A.12)$$

and

$$\left[\bar{G}_{ij}^{sym}\right] = \begin{bmatrix} 0 & 0 & 0 & \eta^s\bar{c}_3 & -\eta^s\bar{s}_3 & -g_2\gamma\bar{c}_5 & -g_2\gamma\bar{s}_5 & -g_3\gamma\bar{c}_5 & -g_3\gamma\bar{s}_5 \\ 0 & 0 & 0 & g_1\bar{s}_3 & g_1\bar{c}_3 & g_5\bar{s}_5 & g_5\bar{c}_5 & g_6\bar{s}_5 & g_6\bar{c}_5 \\ 0 & 0 & 0 & g_4\bar{s}_3 & g_4\bar{c}_3 & g_6\bar{s}_5 & g_6\bar{c}_5 & g_7\bar{s}_5 & g_7\bar{c}_5 \\ \bar{c}_1 & 0 & 0 & \bar{s}_2 & \bar{c}_2 & 0 & 0 & 0 & 0 \\ a_3\bar{c}_1 & a_6\bar{c}_4 & a_8\bar{c}_4 & g_1\bar{s}_2 & g_1\bar{c}_2 & g_5\bar{s}_4 & g_5\bar{c}_4 & g_6\bar{s}_4 & g_6\bar{c}_4 \\ a_5\bar{c}_1 & a_8\bar{c}_4 & a_{11}\bar{c}_4 & g_4\bar{s}_2 & g_4\bar{c}_2 & g_6\bar{s}_4 & g_6\bar{c}_4 & g_7\bar{s}_4 & g_7\bar{c}_4 \\ \eta\bar{s}_1 & a_3\gamma\bar{s}_4 & a_5\gamma\bar{s}_4 & -g_8\eta^s\bar{c}_2 & g_8\eta^s\bar{s}_2 & g_1\gamma\bar{c}_4 & g_1\gamma\bar{s}_4 & g_4\gamma\bar{c}_4 & g_4\gamma\bar{s}_4 \\ 0 & \bar{s}_4 & 0 & 0 & 0 & \bar{c}_4 & \bar{s}_4 & 0 & 0 \\ 0 & 0 & \bar{s}_4 & 0 & 0 & 0 & 0 & \bar{c}_4 & \bar{s}_4 \end{bmatrix} \quad (A.13)$$

where

$$(\eta^s)^2 = r_\rho\Omega^2/g_8 - \gamma^2, \quad \bar{s}_1 = \sin\left(\frac{\pi}{2}\eta\right), \quad \bar{s}_2 = \sin\left(\frac{\pi}{2}\eta^s\right), \quad \bar{s}_3 = \sin\left[\frac{\pi}{2}\eta^s(1+r_h)\right],$$

$$\bar{s}_4 = \sinh\left(\frac{\pi}{2}\gamma\right), \quad \bar{s}_5 = \sinh\left[\frac{\pi}{2}\gamma(1+r_h)\right], \quad \bar{c}_1 = \cos\left(\frac{\pi}{2}\eta\right), \quad \bar{c}_2 = \cos\left(\frac{\pi}{2}\eta^s\right),$$

$$\bar{c}_3 = \cos\left[\frac{\pi}{2}\eta^s(1+r_h)\right], \quad \bar{c}_4 = \cosh\left(\frac{\pi}{2}\gamma\right), \quad \bar{c}_5 = \cosh\left[\frac{\pi}{2}\gamma(1+r_h)\right], \quad (A.14)$$

$$g_1 = \frac{e_{15}}{c_{44}^*}m_1^s, \quad g_2 = \frac{e_{15}}{c_{44}^{*s}}m_1^s, \quad g_3 = \frac{h_{15}}{c_{44}^{*s}}m_2^s, \quad g_4 = \frac{h_{15}}{c_{44}^*}m_2^s,$$

$$g_5 = \frac{e_{15}^2}{c_{44}^*}b_1^s, \quad g_6 = \frac{e_{15}h_{15}}{c_{44}^*}b_2^s, \quad g_7 = \frac{h_{15}^2}{c_{44}^*}b_3^s, \quad g_8 = \frac{c_{44}^{*s}}{c_{44}^*}$$

In (A.12) and (A.13), $\eta$ and $a_3, a_5, a_6, a_8, a_{11}$ are defined by (34) and (39), respectively.

After expanding the trigonometric functions $\bar{s}_3, \bar{c}_3$ and the hyperbolic functions $\bar{s}_5, \bar{c}_5$ in (A.12) and (A.13), taking the limit $\gamma \to \infty$, and making use of (51) and the basic properties of the determinant, (A.12) and (A.13) become

$$\begin{vmatrix} \eta^s \hat{s}_2 & g_2\gamma \hat{c}_4 & g_3\gamma \hat{c}_4 & \eta^s \hat{c}_2 & g_2\gamma \hat{s}_4 & g_3\gamma \hat{s}_4 \\ g_1 \hat{c}_2 & g_5 \hat{s}_4 & g_6 \hat{s}_4 & -g_1 \hat{s}_2 & g_5 \hat{c}_4 & g_6 \hat{c}_4 \\ g_4 \hat{c}_2 & g_6 \hat{s}_4 & g_7 \hat{s}_4 & -g_4 \hat{s}_2 & g_6 \hat{c}_4 & g_7 \hat{c}_4 \\ a_3 - g_1 & a_6 & a_8 & 0 & -g_5 & -g_6 \\ a_5 - g_4 & a_8 & a_{11} & 0 & -g_6 & -g_7 \\ -\eta' & (a_3 - g_1)\gamma & (a_5 - g_4)\gamma & -g_8 \eta^s & 0 & 0 \end{vmatrix} = 0, \qquad (A.15)$$

where

$$\hat{s}_2 = \sin\left(\frac{\pi}{2}\eta^s r_h\right), \quad \hat{s}_4 = \sinh\left(\frac{\pi}{2}\gamma r_h\right), \quad \hat{c}_2 = \cos\left(\frac{\pi}{2}\eta^s r_h\right), \quad \hat{c}_4 = \cosh\left(\frac{\pi}{2}\gamma r_h\right) \quad (A.16)$$

Actually, (A.15) represents the exact dispersion relation for the SH surface wave in the magneto-electro-elastic half-space covered by a surface magneto-electro-elastic layer.

It should be noted here that the exact SH wave dispersion relations for three-layer magneto-electro-elastic plates given in this appendix are valid not only for nano-plates, but also for general three-layer plates at macro-scales.